\documentclass[twocolumn]{aastex631}

\newcommand{\nustar}{\textit{NuSTAR}}

\newcommand{\nicer}{\textit{NICER}}
\newcommand{\ixpe}{\textit{IXPE}}
\newcommand{\integral}{\textit{INTEGRAL}}
\newcommand{\src}{Cygnus~X-2}

\usepackage{amsmath}
\usepackage{graphicx}
\usepackage{parskip}
\usepackage{float}

\begin{document}

\title{Decoding Cygnus X-2: The Critical Role of Reflection in IXPE Data}

\author[0000-0003-2845-1009]{Honghui Liu}
\affiliation{Center for Astronomy and Astrophysics, Center for Field Theory and Particle Physics and Department of Physics, Fudan University, 200438 Shanghai, China.}
\affiliation{Institut f\"ur Astronomie und Astrophysik, Eberhard-Karls Universit\"at T\"ubingen, D-72076 T\"ubingen, Germany}

\author[0000-0002-9639-4352]{Jiachen Jiang}
\affiliation{Department of Physics, University of Warwick, Gibbet Hill Road, Coventry CV4 7AL, UK}

\author[0000-0002-5311-9078]{Adam Ingram}
\affiliation{School of Mathematics, Statistics, and Physics, Newcastle University, Newcastle upon Tyne NE1 7RU, UK}

\author[0000-0002-3180-9502]{Cosimo Bambi}
\altaffiliation{bambi@fudan.edu.cn}
\affiliation{Center for Astronomy and Astrophysics, Center for Field Theory and Particle Physics and Department of Physics, Fudan University, 200438 Shanghai, China.}
\affiliation{School of Humanities and Natural Sciences, New Uzbekistan University, Tashkent 100001, Uzbekistan}

\author{Andrew C. Fabian}
\affiliation{Institute of Astronomy, University of Cambridge, Madingley Road, Cambridge CB3 0HA, UK}

\author{Ruben Farinelli}
\affiliation{INAF - Osservatorio di Astrofisica e Scienza dello Spazio di Bologna, Via P. Gobetti 101, I-40129 Bologna, Italy}

\author{Renee Ludlam}
\affiliation{Department of Physics and Astronomy, Wayne State University, 666 W. Hancock St., 48201, Detroit, MI, USA}

\author{Nathalie Degenaar}
\affiliation{Anton Pannekoek Institute for Astronomy, University of Amsterdam, Postbus 94249, NL-1090 GE Amsterdam, the Netherlands}

\author{Jakub Podgorny}
\affiliation{Astronomical Institute of the Czech Academy of Sciences, Bo\v{c}n\'{i} II 1401/1, 14100 Praha 4, Czech Republic}

\author{Andrea Santangelo}
\affiliation{Institut f\"ur Astronomie und Astrophysik, Eberhard-Karls Universit\"at T\"ubingen, D-72076 T\"ubingen, Germany}

\author{James F. Steiner}
\affiliation{Harvard-Smithsonian Center for Astrophysics, 60 Garden Street, Cambridge, MA 02138, USA}

\author[0000-0003-3626-9151]{Andrew J. Young}
\affiliation{H.H. Wills Physics Laboratory, Tyndall Avenue, Bristol BS8 1TL, UK}

\author{Zuobin Zhang}
\affiliation{Astrophysics, Department of Physics, University of Oxford, Keble Road, Oxford OX1 3RH, UK}

\begin{abstract}
We present a spectro-polarimetric re-analysis of the first IXPE observation of Cygnus X-2 which we determine to be mainly in the normal branch, from quasi-simultaneous observations with NuSTAR, NICER, and INTEGRAL. We measure the hard X-ray polarization angle and find it to be consistent with the previously measured position angle of the radio jet. Leveraging NuSTAR's detection of both the relativistic Fe K emission line and the Compton hump, we constrain the flux contribution of the reflected emission from the inner accretion disk to be 10\% of the total X-ray flux in the IXPE energy band. Unlike previous studies that modeled only the Fe K emission line, we fit the full-band reflection spectrum using a fully relativistic disk model. There is strong degeneracy between the Comptonized and reflection components. Given that the Comptonized component is not expected to be highly polarized, a polarization degree of approximately 20\% for the reflection component could explain the X-ray polarization data from IXPE. We also discuss the disk inclination angle inferred from our spectro-polarimetric modeling, as well as other possible explanations for the data. 
\end{abstract}


\section{Introduction} \label{sec:intro}

X-ray binaries (XRBs) are systems in which a compact object (black hole or neutron star) is accreting matter from its donor star. The accretion process efficiently converts the gravitational energy of the accreted matter into electromagnetic radiation \citep{Thorne1974}. An optically thick and geometrically thin accretion disk \citep{Shakura1973} is thought to be present, emitting multi-temperature blackbody radiation normally peaked in the soft X-ray band. Additionally, a harder non-thermal component is usually observed, which is believed to originate from inverse-Compton scattering of disk seed photons by a hot plasma \citep{Haardt1993ApJ...413..507H,Done2007} close to the compact object, often referred to as the \textit{corona}. 

A third component, the reprocessed coronal emission by the optically thick accretion disk, is often present \citep[\textit{reflection} component, e.g.,][]{Cackett2010ApJ...720..205C, Ludlam2024Ap&SS.369...16L, Garcia2015, Walton2016}. The reflection component is characterized by a relativistically broadened iron K$_{\alpha}$ line around 6--7 keV and a Compton hump peaked near 30 keV \citep{Fabian1989, George1991}. Analyzing the reflection component is a powerful tool for understanding the accretion geometry \citep[e.g.,][]{Cackett2008ApJ...674..415C, Jiang2019gx339, Liu2022, Liu2023ApJ...950....5L, Mondal2020MNRAS.494.3177M, Ludlam2022ApJ...927..112L}, compact objects \citep[e.g., spins of black holes,][]{Reynolds2021, Bambi2021, Miller2013ApJ...779L...2M}, and gravity theories \citep[e.g.,][]{Bambi2017, Liu2019}.


For weakly magnetized neutron star (NS) low-mass X-ray binaries (LMXBs), the accretion flow from the companion star is stopped by the surface of the NS. This causes the formation of a boundary layer (BL) between the accretion disk and the NS \citep{Popham2001ApJ...547..355P}. In addition, a spreading layer (SL) that partially covers the surface of the NS can also be formed \citep{Inogamov1999AstL...25..269I}. The SL can extend to higher latitudes if the accretion rate increases (see Fig.~2 of \citealt{Farinelli2024A&A...684A..62F}). The BL/SL could be the origin of the non-thermal Comptonization emission in NS XRBs \citep{Revnivtsev2013MNRAS.434.2355R}.

NS LMXBs are known to be highly variable sources and their variability can be traced on the X-ray color-color diagram (CCD) or the hardness-intensity diagram (HID). Based on the shapes of their tracks on the CCDs, NS LMXBs can broadly be divided into two categories: Z sources and atoll sources \citep{Hasinger1989A&A...225...79H}. The main difference between the two types of sources is the mass accretion rate, with the Z sources accreting near the Eddington limit ($L_{\rm Edd}$) and atoll sources accreting at a lower accretion rate (e.g., 0.001--0.5 $L_{\rm Edd}$). The Z-track can be divided into three branches: the horizontal branch (HB), the normal branch (NB), and the flaring branch (FB). The track of atoll sources can be divided into the lower part banana state and the upper part island state \citep[see][for a recent review]{DiSalvo2023arXiv231112516D}. These systems are among the brightest sources of the X-ray sky and are important for studies of accretion physics and fundamental physics (e.g., the equation of state of ultra-dense matter, \citealt{Degenaar2018ASSL..457..185D}).


Over the last decades, X-ray spectroscopy and timing techniques have been leading the study of NS LMXBs. However, the accretion geometry still remains unclear. The launch of the Imaging X-ray Polarimetry Explorer (\textit{IXPE}, \citealt{Soffitta2021AJ....162..208S, Weisskopf2022JATIS...8b6002W}) at the end of 2021 reopened the polarization window and has the potential to break degeneracies in traditional methods, such as determining the geometry of emission regions \citep[e.g.,][]{Gnarini2022MNRAS.514.2561G, 2022Sci...378..650K, Veledina2024NatAs...8.1031V}. It can also provide additional constraints on black hole spins \citep[e.g.,][]{Dovvciak2008MNRAS.391...32D, Schnittman2009ApJ...701.1175S, Marra2024A&A...684A..95M} and other strong gravity effects \citep[e.g.,][]{Dovvciak2004MNRAS.355.1005D, Krawczynski2012ApJ...754..133K, Liu2015EPJC...75..383L,Steiner2024ApJ...969L..30S}.


In this work, we will study the spectral and polarization properties of Cygnus X-2. The source is a Z-type neutron star low-mass XRB located at a distance of $11.3_{-0.8}^{+0.9}$ kpc \citep{Ding2021PASA...38...48D}. The orbital inclination of the system is $62.5^{\circ} \pm 4^{\circ}$ \citep{Orosz1999MNRAS.305..132O}. Previous spectral studies have shown that the source exhibits clear reflection features \citep[e.g.,][]{Mondal2018MNRAS.474.2064M, Ludlam2022ApJ...927..112L}. However, measurements of its disk inclination angle using the reflection method are not entirely consistent, with some being close to the binary inclination \citep{Ludlam2022ApJ...927..112L}, while others favor a lower inclination \citep{Shaposhnikov2009ApJ...699.1223S, Mondal2018MNRAS.474.2064M}. The source can be highly variable, transitioning between branches on a short timescale of hours \citep[e.g.][]{Piraino2002ApJ...567.1091P, Sudha2025ApJ...978...75S, Zhang2025ApJ...987..107Z}. On longer timescales (e.g., weeks to months), the morphology and position of its Z-track on the CCD/HID can also change \citep{Kuulkers1996A&A...311..197K}.

\src{} was observed by \ixpe{} a few months after its launch when the source was likely on the horizontal branch (HB) or normal branch (NB) of the Z-track \citep{Farinelli2023MNRAS.519.3681F}. Theoretical modelling of the boundary layer polarization by \cite{Farinelli2024A&A...684A..62F} has shown that its polarization degree (PD) for typical optical depths $\tau \ga 4$ never exceeds 1\%. Calculations by \cite{Bobrikova2025A&A...696A.181B} show that the PD of the spreading layer does not exceed 1.5\%, regardless of its geometry. Therefore the observed source PD in the IXPE band of about 1.7\% requires an additional component that \cite{Farinelli2023MNRAS.519.3681F} identified as accretion disk reflection. However, this is yet to be validated from detailed spectro-polarimetric analysis including a full relativistic reflection model. 

The aim of this paper is to further test whether the reflection component significantly contributes to the polarization properties of \src{}. The manuscript is structured as follows: In Sec.~\ref{sec:data_reduction}, we describe the observations and data reduction. The analysis of the data is presented in Sec.~\ref{sec:analysis}. Sec.~\ref{sec:discussion} is devoted to the discussion of the results.

\section{Observations and Data reduction} \label{sec:data_reduction}

\begin{figure*}[htbp]
    \centering
    \includegraphics[width=0.45\textwidth]{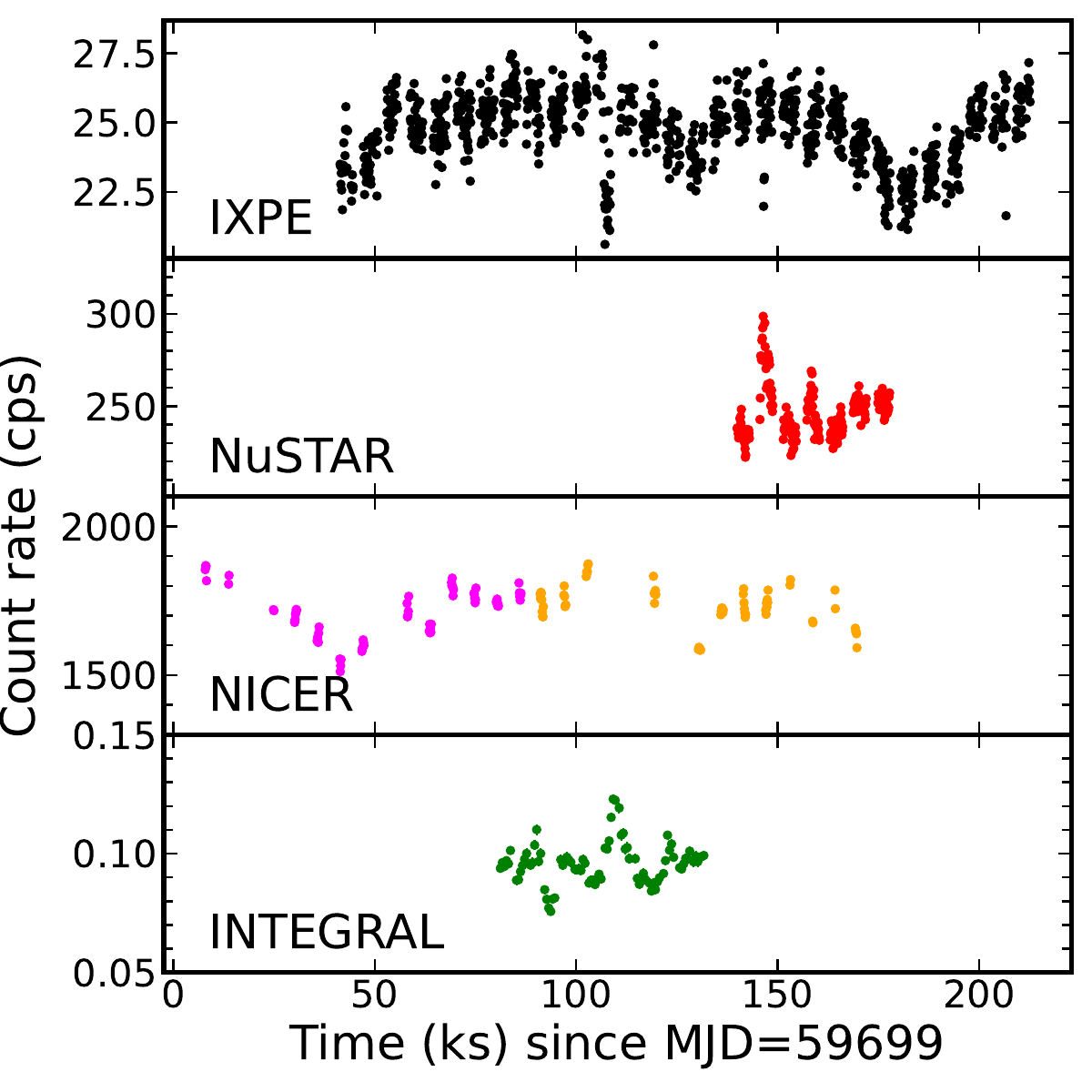}
    \includegraphics[width=0.45\textwidth]{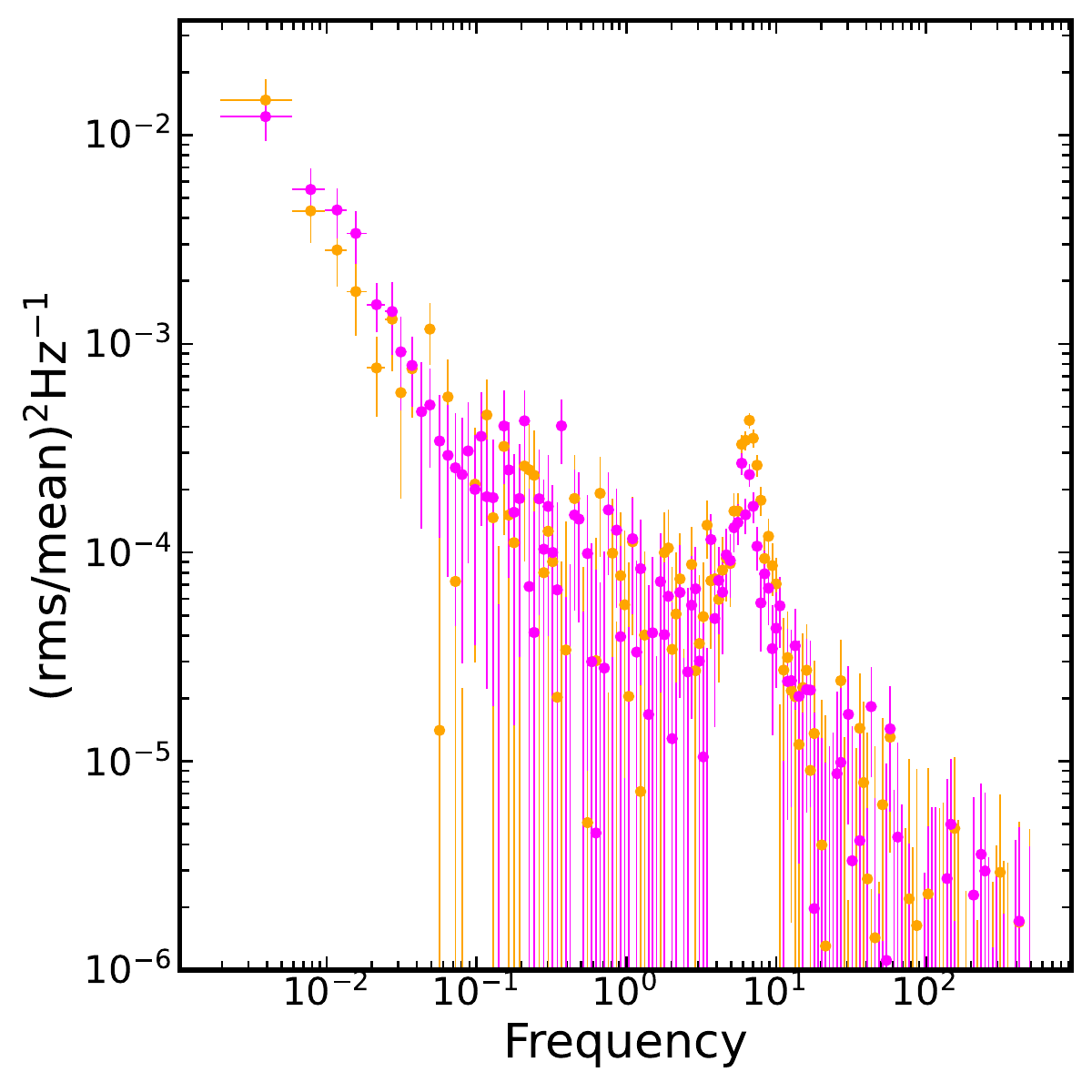}
    \caption{
    (Left) Light curves of \src{} from \textit{IXPE} (DU1), \textit{NuSTAR} (FPMA), \textit{NICER} and \textit{INTEGRAL} (JEM-X1). The consecutive observations of \textit{IXPE} and \textit{NICER} are distinguished using different colors. (Right) Power spectral density for the two \nicer{} observations.
    }
    \label{lightcurve}
\end{figure*}

In this study, we analyze the \textit{IXPE} observation of \src{} performed on 30 April 2022 \citep{Farinelli2023MNRAS.519.3681F}. Following \cite{Farinelli2023MNRAS.519.3681F}, we include quasi-simultaneous \textit{NICER} and \textit{INTEGRAL} observations to constrain the broadband energy spectrum. Additionally, data from a contemporary \textit{NuSTAR} observation (PI: J. Jiang) is included to better constrain the spectral components. Details of observations used in this work are listed in Table \ref{obs}. Light curves of the source from these observations are shown in Fig.~\ref{lightcurve}.

\subsection{\textit{IXPE}}

The \textit{IXPE} observations are reduced using \texttt{ixpeobssim} v31.1.1 \citep{Baldini2022} and CALDB version 20211209. We download level 2 data from the \textit{IXPE} archive\footnote{\url{https://heasarc.gsfc.nasa.gov/docs/ixpe/archive/}}. The source event files for each of the three detector units (DUs) are extracted from 100 arcsec-radius circular regions using \texttt{xpselect}. Given that the source is bright, we neglect the background \citep{DiMarco2023AJ....165..143D}. The \textit{I}, \textit{Q}, and \textit{U} spectra are then produced using the `weighted' \texttt{pha} algorithm of the \texttt{xpbin} task \citep{DiMarco2022AJ....163..170D}. The response file of the effective area for each spectrum is generated using the \texttt{ixpecalcarf} task. We use \textit{IXPE} data in the 2--8 keV band for DU1, DU2 and DU3. The Stokes \textit{I} spectra are binned to have a minimum counts of 20 per bin, while the Stokes \textit{Q} and \textit{U} spectra are binned to have a constant bin width of 0.2 keV.

\subsection{\textit{NuSTAR}}

Cleaned event files for both FPMA and FPMB onboard \textit{NuSTAR} \citep{nustar} are produced using the tool \texttt{nupipeline} v0.4.9 and the calibration database (CALDB) version 20230307. Due to the high count rate, we implemented the bright source flag \texttt{statusexpr="(STATUS==b0000xxx00xxxx000) \&\& (SHIELD==0)"} for \texttt{nupipeline}. The source spectra are extracted from a circular region centered on the source with a radius of 200 arcsec, while the background spectra are extracted from source-free areas using polygon regions. We fit the spectra in the 3--40 keV range due to the high background above 40 keV. Spectra are grouped using \texttt{ftgrouppha}\footnote{\url{https://heasarc.gsfc.nasa.gov/lheasoft/help/ftgrouppha.html}} with the options ``\texttt{grouptype=optmin groupscale=20}''. This applies the ``optimal binning'' strategy \citep{Kaastra2016A&A...587A.151K} while also ensuring a minimal of 20 counts per bin.

\subsection{\textit{NICER}}

\textit{NICER} \citep{nicer} data are first processed using the \textit{nicerl2} task with CALDB v20240206 and geomagnetic data as of March 2024\footnote{\url{https://heasarc.gsfc.nasa.gov/docs/nicer/analysis_threads/geomag/}}. The spectra and light curves are then extracted using \texttt{nicerl3-spect} and \texttt{nicerl3-lc} without binning the spectra. The default background model (\texttt{SCORPEON}\footnote{\url{https://heasarc.gsfc.nasa.gov/lheasoft/ftools/headas/niscorpeon.html}}) is used to produce the background spectral files. Spectra from the two observations are fitted simultaneously in the 1--10 keV band after rebinning with the same strategy as the \nustar{} data.

\subsection{\textit{INTEGRAL}}

\textit{INTEGRAL} observations of \src{} from April 30 to May 01 2022 are processed in the same way as described in \cite{Farinelli2023MNRAS.519.3681F}. We obtain two spectra from the two identical detectors of JEM-X and one spectrum from ISGRI.

\subsection{Lightcurves and tracks}

\begin{figure*}[htbp]
    \centering
    \includegraphics[width=0.95\textwidth]{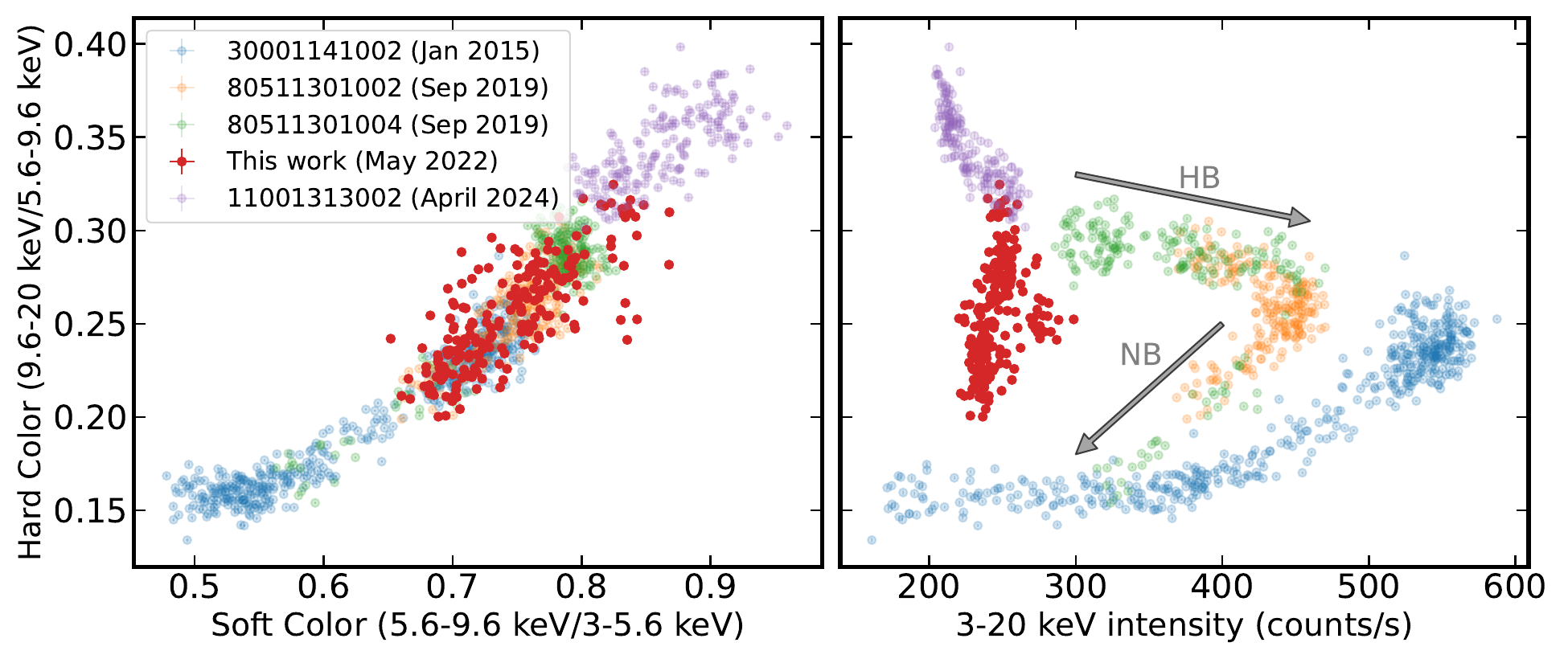}
    \caption{
    The color-color diagram (left) and hardness-intensity diagram (right) plotted with the 100-s binned \nustar/FPMA lightcurves. All avaliable \nustar{} data are included. The observation analyzed in this work is shown in red. Other observations are color-coded based on their observation IDs.
    }
    \label{ccd}
\end{figure*}

We present lightcurves from the observations in Fig.~\ref{lightcurve}, where short flares and dips are occasionally observed. We find no type-I burst after visually inspecting the lightcurve with a time resolution of 1~s. In Fig.~\ref{ccd}, we show the track of the source on the CCD and HID, including all available archival \nustar{} data. The two 2019 observations trace a Z-track in the HID, clearly showing the horizontal and normal branches. However, the branches are less distinguishable in the CCD. The other observations show more complex tracks that do not overlap with the 2019 track on the HID. Such secular variations of the Z-track are known for \src{} \citep[e.g.,][]{Kuulkers1996A&A...311..197K}. The observations analyzed in this work appears to be mainly in the normal branch, and could cover periods of the flaring branch. This is further supported by the timing properties of the overlapping \nicer{} observation (see below). 

We still note that the \nustar{} observation analyzed here occupies a small region in both the CCD and HID, indicating little variability in flux or spectral shape. It is known the branches of NS LMXBs are associated with characteristic timing properties in their power spectral density (PSD, see Fig. 2 of \citealt{DiSalvo2023arXiv231112516D}). Therefore, we calculate the PSDs of the two \nicer{} observations, which overlap with the \nustar{} observation and cover half of the \ixpe{} observations (see the right panel of Fig.~\ref{lightcurve}). The two PSDs show remarkable similarity, indicating little variability in the system between these observations. 
The PSDs are characterized by a power-law-like continuum and a quasi-periodic oscillation at around 7~Hz, which are typical for the normal branch \citep{vanderKlis1987ApJ...316..411V, Hasinger1989A&A...225...79H}.

\begin{table*}
    \centering
    \caption{Summary of observations analyzed in this work}
    \label{obs}
    \renewcommand\arraystretch{1.5}
    \begin{tabular}{lccc}
        \hline\hline
        Instrument & Obs Date & Obs ID & Exposure (ks) \\
        \hline
        \textit{NuSTAR} & 2022-05-01 & 30801012002 & 15.2 \\
        \hline
        \textit{NICER} & 2022-04-30 & 5034150102 & 3.6 \\
                       & 2022-05-01 & 5034150103 & 4.5 \\
        \hline
        \textit{IXPE} & 2022-04-30 & 01001601 & 93.2 \\
        \hline
        \textit{INTEGRAL} & 2022-04-30 & 19700060001 & 50 \\
        \hline
    \end{tabular} \\
\end{table*}

\section{Spectro-polarimetric analysis}
\label{sec:analysis}

Spectral analysis (including the Stokes \textit{Q} and \textit{U} spectra) is conducted using \texttt{XSPEC} v12.13.0c \citep{xspec}. The element abundances from \cite{Wilms2000} and the cross-sections from \cite{Verner1996} are implemented throughout the analysis. $\chi^2$ statistics is used to find the best-fit values and uncertainties.

\subsection{2--8 keV polarization properties}

As the first step, we use two different methods to estimate the 2--8 keV band polarization properties: (a) the \texttt{pcube} algorithm of \texttt{ixpeobssim} and (b) fitting the \textit{I}, \textit{Q}, and \textit{U} spectra in \texttt{XSPEC}.

\subsubsection{\texttt{PCUBE}}

In the first method, the Stokes parameters are summed from events of the three DUs in the source regions \citep{Kislat2015APh....68...45K, Baldini2022}.
We find a model-independent PD of $p = 1.4 \pm 0.3\%$ and a PA of $\psi = 132^{\circ} \pm 6^{\circ}$ (1-$\sigma$). In comparison, the minimum detectable polarization at 99\% significance (MDP$_{99}$) for the analyzed observations is 0.8\% in the 2--8 keV band. The PD and PA we measured are consistent with what found by \cite{Farinelli2023MNRAS.519.3681F} ($p = 1.85 \pm 0.29\%$, $\psi = 140^{\circ} \pm 4^{\circ}$).

\subsubsection{\texttt{XSPEC fitting}}

Secondly, following \cite{Farinelli2023MNRAS.519.3681F}, we fit all spectra from \nicer{}, \nustar{}, \integral{} and \ixpe{} simultaneously with the model \texttt{polconst*mbpo*tbabs*(diskbb+comptt+gaussian)} (Model 1). The \texttt{tbabs} \citep{Wilms2000} component accounts for absorption by the interstellar medium. The model includes the thermal emission from the disk \citep[\texttt{diskbb},][]{Mitsuda1984}, the Comptonization component from the boundary layer \citep[\texttt{comptt},][]{Titarchuk1994ApJ...434..570T} and the iron emission line (\texttt{gaussian}). 
we use the \texttt{mbpo} \citep{Ingram2017MNRAS.464.2979I} model to account for cross-calibration uncertainties between \ixpe{}, \nustar{} and \nicer{}. The \texttt{mbpo} component multiplies the total model by a constant (Norm), as well as a broken power-law with the index $\Delta\Gamma_{1}$ for $E<E_{\rm br}$ and $\Delta\Gamma_{2}$ for $E>E_{\rm br}$. In this work, we find that a single-index power-law is sufficient to fit the data, i.e., $\Delta\Gamma_{1}=\Delta\Gamma_{2}$.
We also apply a gain shift to the \textit{IXPE} data using the \texttt{gain fit} command in \texttt{XSPEC}.
Finally, \texttt{polconst} adds a constant (i.e., energy-independent) PD and PA to the model it multiplies. 

\begin{figure*}[htbp]
    \centering
    \includegraphics[width=0.45\textwidth]{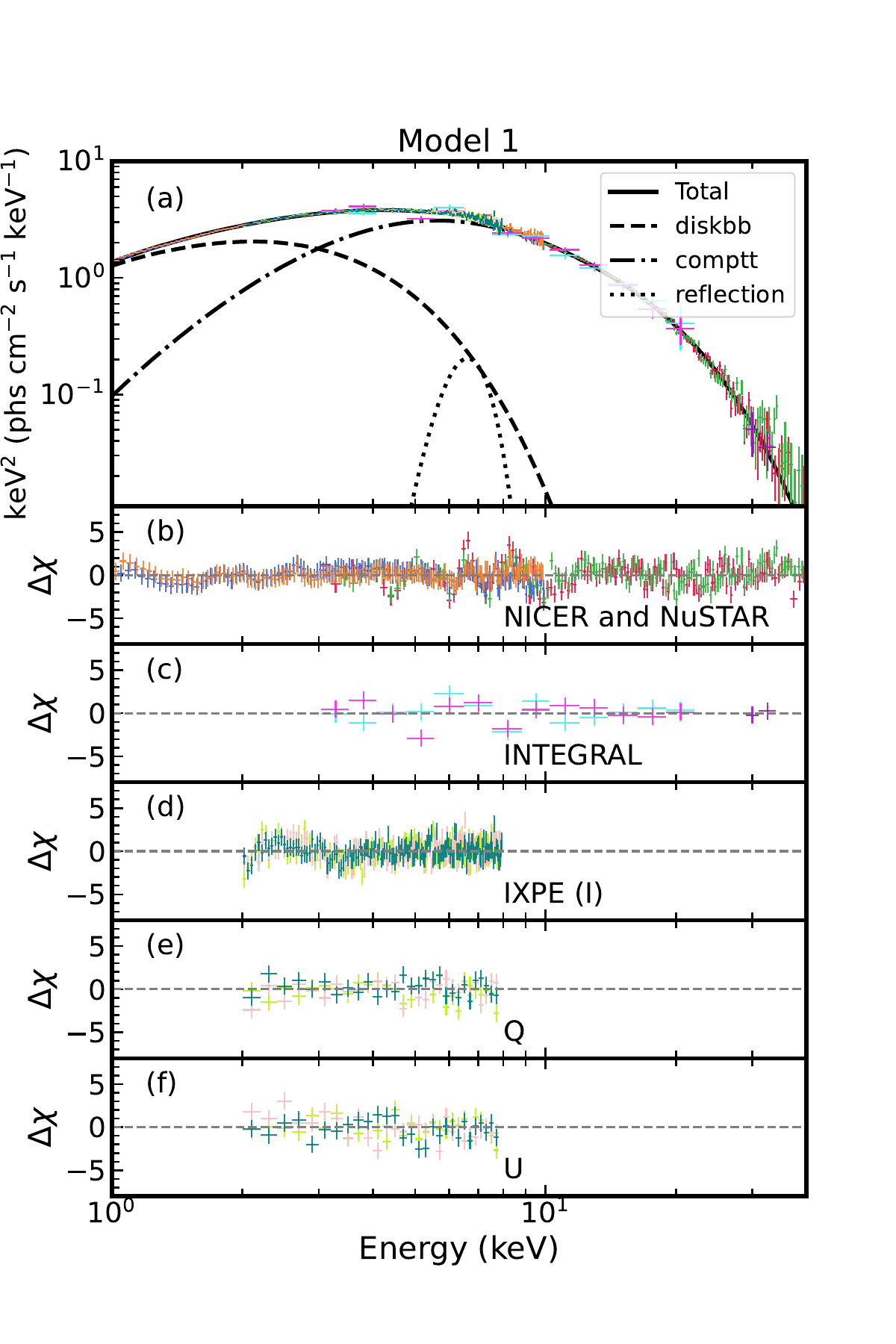}
    \includegraphics[width=0.45\textwidth]{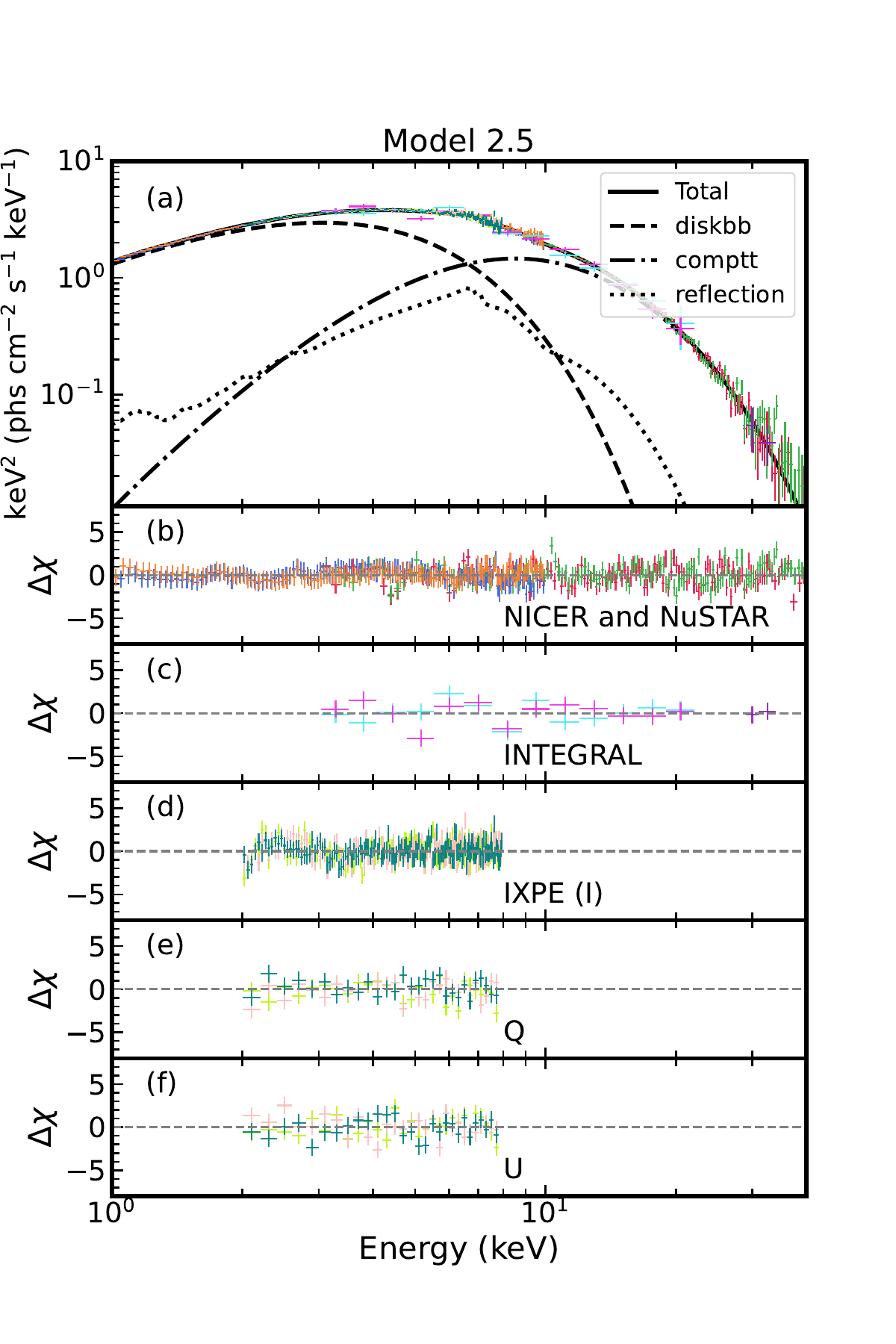}
    \caption{
    The best-fit model components and residuals for Model 1 (left) and Model 2.5 (right). Colors are coded to refer: (b) \textit{NuSTAR}-FPMA (red), \textit{NuSTAR}-FPMB (green), the \textit{NICER} observation on April 30 (blue) and May 01 (orange); (c) \textit{INTEGRAL} ISGRI (purple), JEM-X1 (cyan) and JEM-X2 (pink); (d,e,f) \textit{IXPE} DU1 (yellow), DU2 (light pink) and DU3 (teal).
    }
    \label{eemod_delchis}
\end{figure*}

\begin{figure}[htbp]
    \centering
    \includegraphics[width=0.45\textwidth]{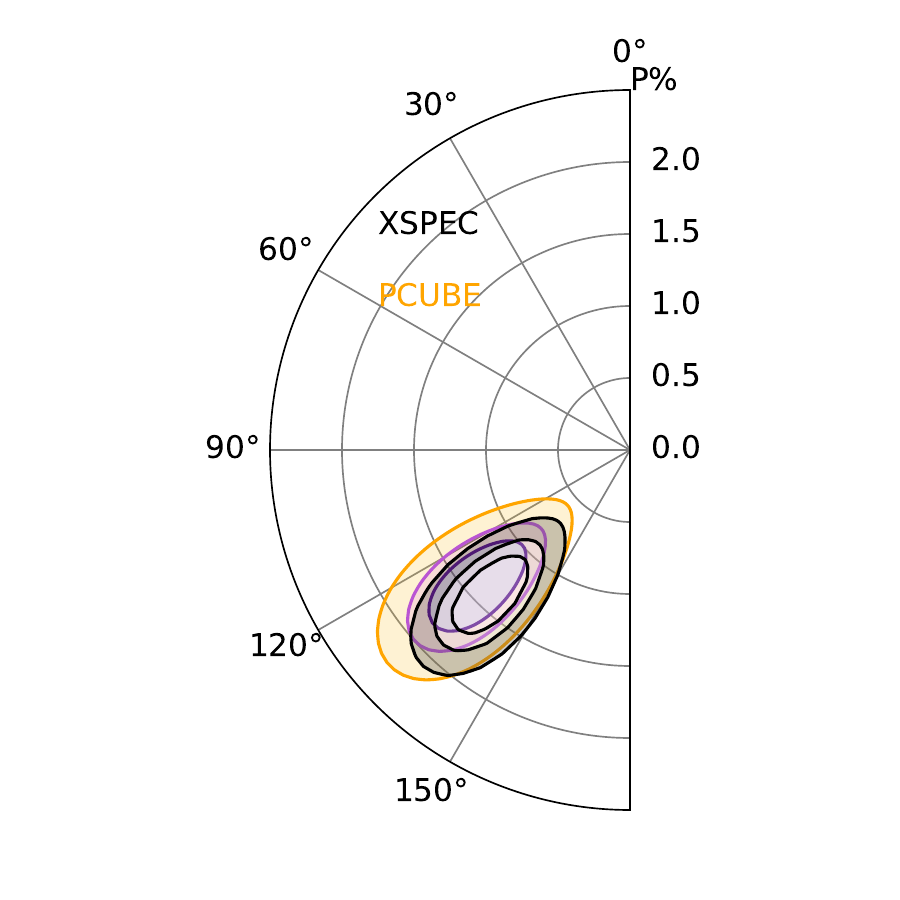}
    \caption{
    Constraints on the PA and PD in the 2--8~keV band. Contours are shown at the 68\%, 90\% and 99\% confidence levels. Black contours represent the result of our spectro-polarimetric analysis using \texttt{XSPEC}. The orange-magenta-violet contours are for the \texttt{pcube} algorithm.
    }
    \label{pcube_vs_xspec}
\end{figure}

The model (Model 1) provides a good fit to the data, with $\chi^2/\nu=1281.2/1180$. The best-fit parameters are listed in Tab.~\ref{best-fit}. The best-fit spectral components and residuals are shown in Fig.~\ref{eemod_delchis}. We obtain a PD of $p=1.4\pm 0.2\%$ and a PA of $\psi=136^{\circ}\pm 5^{\circ}$ (1-$\sigma$), which are consistent with the model-independent \texttt{pcube} algorithm. Results of the two methods are also reported in Fig.~\ref{pcube_vs_xspec}, which are in agreement with the result of Fig. 5 of \cite{Farinelli2023MNRAS.519.3681F}. Note that the measured PA is consistent with the direction of the radio jet \citep{Spencer2013MNRAS.435L..48S}.

\cite{Farinelli2023MNRAS.519.3681F} applied the \texttt{polconst} model to every spectral component and found that the PA of the disk thermal component is perpendicular to the total PA in the 2--8 keV band. The Comptonization component from the boundary layer or spreading layer is likely responsible for the overall X-ray polarization properties, but a strong contribution from the reflection component cannot be excluded. We repeat the analysis of \cite{Farinelli2023MNRAS.519.3681F} with the model: \texttt{mbpo * tbabs * (polconst * diskbb + polconst * comptt + polconst * gaussian)}. The PD of the \texttt{gaussian} component is fixed to 0 becasue the fluoresent process is supposed to be isotropic \citep{Churazov2002MNRAS.330..817C, Veledina2024NatAs...8.1031V}. The PA of the \texttt{diskbb} component (PA$_{\rm disk}$) is linked to that of the \texttt{comptt} component (PA$_{\rm comp}$) by the relation: PA$_{\rm disk}$=PA$_{\rm comp}-90^{\circ}$. The spectral parameters reported in Tab.~\ref{best-fit} remain unchanged and the $\chi^2/\nu=1269.9/1179$. We obtain PA$_{\rm disk}=47\pm 7 ^{\circ}$, PD$_{\rm disk}<2.3\%$ and PD$_{\rm comp}=3\pm 1\%$. These measurements are consistent with those obtained by \cite{Farinelli2023MNRAS.519.3681F}. Similar PDs of the Comptonization component have been found in other Z sources \citep[e.g.,][]{LaMonaca2024A&A...691A.253L,LaMonaca2025A&A...702A..40L, LaMonaca2025A&A...702A.101L, Lavanya2025ApJ...985..229L}.

\subsection{Implementation of full reflection model}
\label{reflection}

Rather than a simplified approximation of a reflection spectrum as merely a broad Fe-K fluorescent line (e.g., modeled via a gaussian), it is important to consider the full reflection continuum which includes Fe-K emission and back-scattered Compton hump. This, in turn, can also affect the fitted disk and Comptonization terms \citep[e.g.,][]{LaMonaca2024A&A...691A.253L, LaMonaca2025A&A...702A..40L} and change the interpretation of the X-ray polarization measurements. Moreover, the reflection component has its own contribution to the polarization properties, i.e., a combination of the depolarized fluorescent lines and highly polarized continuum \citep{Matt1993MNRAS.260..663M, Poutanen1996MNRAS.283..892P, Podgorny2022MNRAS.510.4723P, Podgorny2025arXiv250723687P}. In the following analysis, we consider these effects to understand the X-ray polarization properties of \src{}.

\begin{figure}[htbp]
    \centering
    \includegraphics[width=0.45\textwidth]{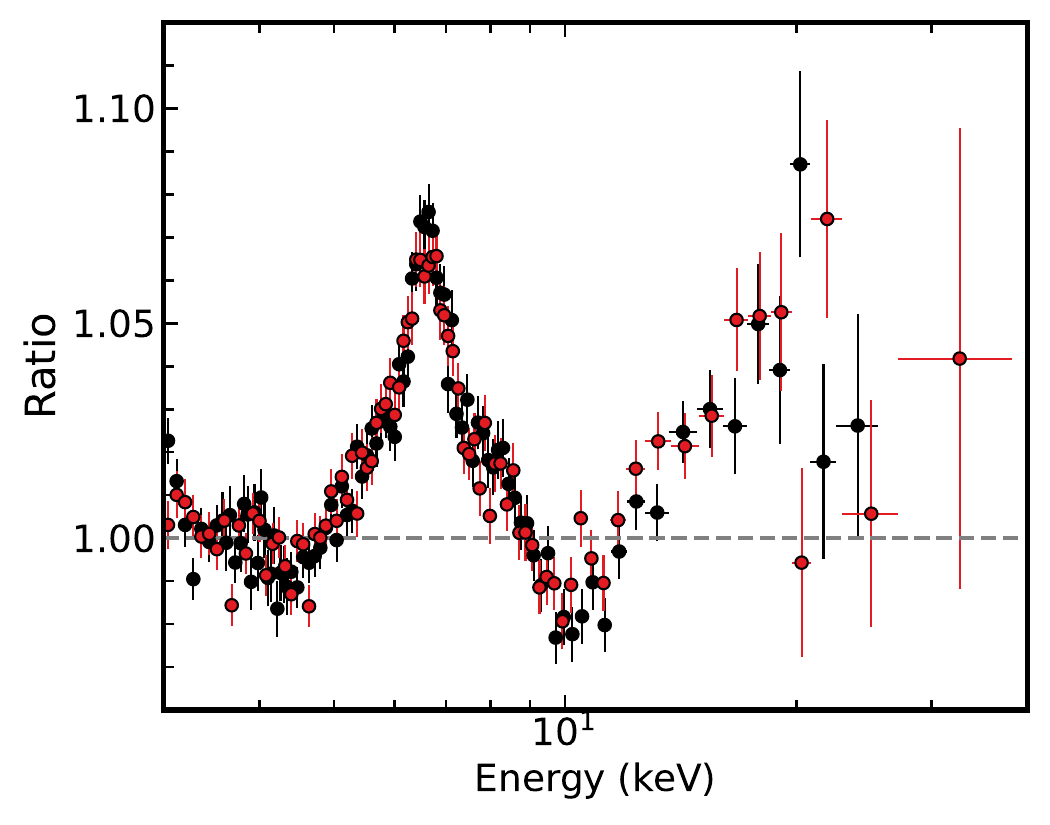}
    \caption{
    Reflection features in the \nustar{} data of \src{}. Data are fitted to a simple absorbed continuum (see the text) in the 3--4 keV and 8--10 keV and 20--40 keV bands, ignoring the bands where the reflection component is strong. Black and red colors represent data from FPMA and FPMB, respectively. The plot is for illustration purpose only.
    }
    \label{ironline}
\end{figure}

To show the reflection features in the data, we first fit the \textit{NuSTAR} spectra with a simple absorbed continuum model: \texttt{const * tbabs * (diskbb + comptt)}. The data-to-model ratios are shown in Fig.~\ref{ironline}, where we observe a clear broad iron line and a relatively weak Compton hump.

We first identify the best spectral model to fit the energy spectrum. The \textit{NICER}, \textit{NuSTAR} and \textit{INTEGRAL} energy spectra of \src{} are fitted with the following models:

\begin{itemize}
\item Model 2.1 - \texttt{mbpo * tbabs * (diskbb + nthcomp + relconv * reflionx\_nth)}, where the \texttt{nthcomp} \citep{Zdziarski1996, Zycki1999} model represents the Comptonization component with the seed photon set to a single-temperature blackbody. \texttt{reflionx\_nth} is the non-relativistic reflection model assuming \texttt{nthcomp} as the incident spectrum.\footnote{The model can be downloaded from \url{https://github.com/honghui-liu/reflionx_tables}. The FITS file is \texttt{reflionx\_HD\_nthcomp\_v2.fits}. Note that in this reflection model, the seed photon spectrum for the incident \texttt{nthcomp} is a single-temperature blackbody.} The \texttt{relconv} \citep{Dauser2010} kernel is used to include relativistic effects. In this scenario, the neutron star provides seed photons for a shell-like Comptonization medium (Eastern model, e.g., \citealt{Mitsuda1989PASJ...41...97M}).

\item Model 2.2 - \texttt{mbpo * tbabs * (bbody + nthcomp + relconv * reflionx\_nth)}, where \texttt{bbody} is a single-temperature blackbody component and the seed photon for \texttt{nthcomp} is set to a disk blackbody. In this scenario, the Comptonization medium has a slab-like geometry above the accretion disk (Western model, e.g., \citealt{White1988ApJ...324..363W}).

\item Model 2.3 - \texttt{mbpo * tbabs * (diskbb + comptt + relconv * reflionx\_bb)}. This model is motivated by \cite{Mondal2018MNRAS.474.2064M}, in which it provides a good fit to the \nustar{} data of \src{} in the normal branch. The \texttt{reflionx\_bb} is a local reflection model assuming a single-temperature blackbody, instead of a power-law, as the incident spectrum.\footnote{The model, \texttt{reflionx\_bb.mod}, can be downloaded from the link in the previous footnote. In this reflection model, the disk electron density is not a free parameter and is assumed to be 10$^{15}$~cm$^{-3}$.} This reflection model is more appropriate for this study since the illuminating continuum is quite soft. The temperature of the illuminating blackbody ($kT_{\rm bb}$) is left as a free parameter. A slab geometry is chosen for \texttt{comptt}.
\end{itemize}

In all three models above, the spin parameter is fixed at $a_* = 0.17$ derived from the spin frequency of the neutron star \citep{Wijnands1998ApJ...493L..87W, Braje2000ApJ...531..447B, Mondal2018MNRAS.474.2064M}, and the emissivity profile is set to be a simple power-law with an index $q = 3$ \citep{Wilkins2018MNRAS.475..748W}. We have tested that the emissivity index could not be constrained by the data. Leaving the emissivity index free does not improve the fit and has no impact on other parameters. The free parameters of the accretion disk are: inclination angle ($i$), inner disk radius ($R_{\rm in}$), ionization parameter ($\xi = 4\pi F / n_{\rm e}$), and iron abundance ($A_{\rm Fe}$). For the \texttt{reflionx\_nth} component, the parameters for the illumination spectrum are linked to those of \texttt{nthcomp}.

\begin{figure}[htbp]
    \centering
    \includegraphics[width=0.45\textwidth]{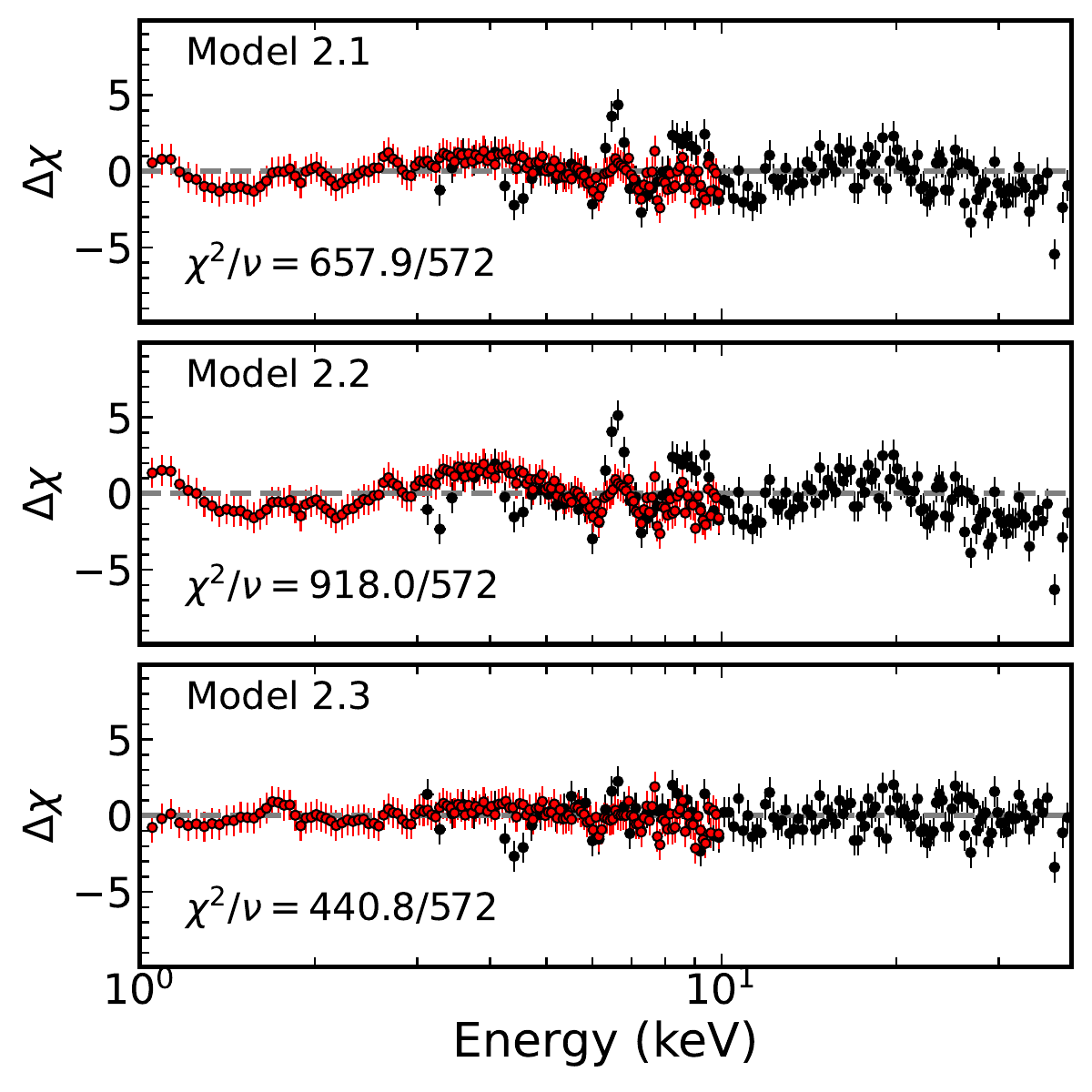}
    \caption{
    Residuals after fitting the energy spectra with the three models described in Sec.~\ref{reflection}. For visual clarity, the data are rebinned, showing only data from \nustar/FPMA (black) and the \nicer{} observation on May 1 (red). See text in Sec.~\ref{reflection} for more details.
    }
    \label{delchis}
\end{figure}

The statistics and residuals for the fitting are shown in Fig.~\ref{delchis}. It is evident that Model 2.3 fits the data best, while the other two models exhibit significant unresolved residuals in the iron band and above 20 keV. Therefore, we proceed with Model 2.3 to investigate the polarization properties of each component.

We include polarization into Model 2.3 by including three instances of \texttt{polconst}, acting on the Comptonization, thermal, and reflection components. The PD of all the three components are free to vary for the fitting. We fit only the PA of the disk component, while the other two PAs are linked to the disk PA in a non-trivial manner. For a standard optically thick accretion disk, the PA of the disk thermal emission (PA$_{\rm disk}$) should be parallel to the disk plane \citep{Chandrasekhar1960ratr.book.....C}. The PA for the reflection component (PA$_{\rm refl}$) should be perpendicular to the plane \citep[e.g., PA$_{\rm refl}-90^{\circ}$ = PA$_{\rm disk}$,][]{Matt1993MNRAS.260..663M}. As for the boundary layer emission, its PA can be either perpendicular or parallel to the disk PA, depending on factors such as optical depth, seed photon distribution, and the exact geometry \citep{Farinelli2024A&A...684A..62F}. Therefore, we consider two configurations for the PAs.

\begin{figure*}[htbp]
    \centering
    \includegraphics[width=0.45\textwidth]{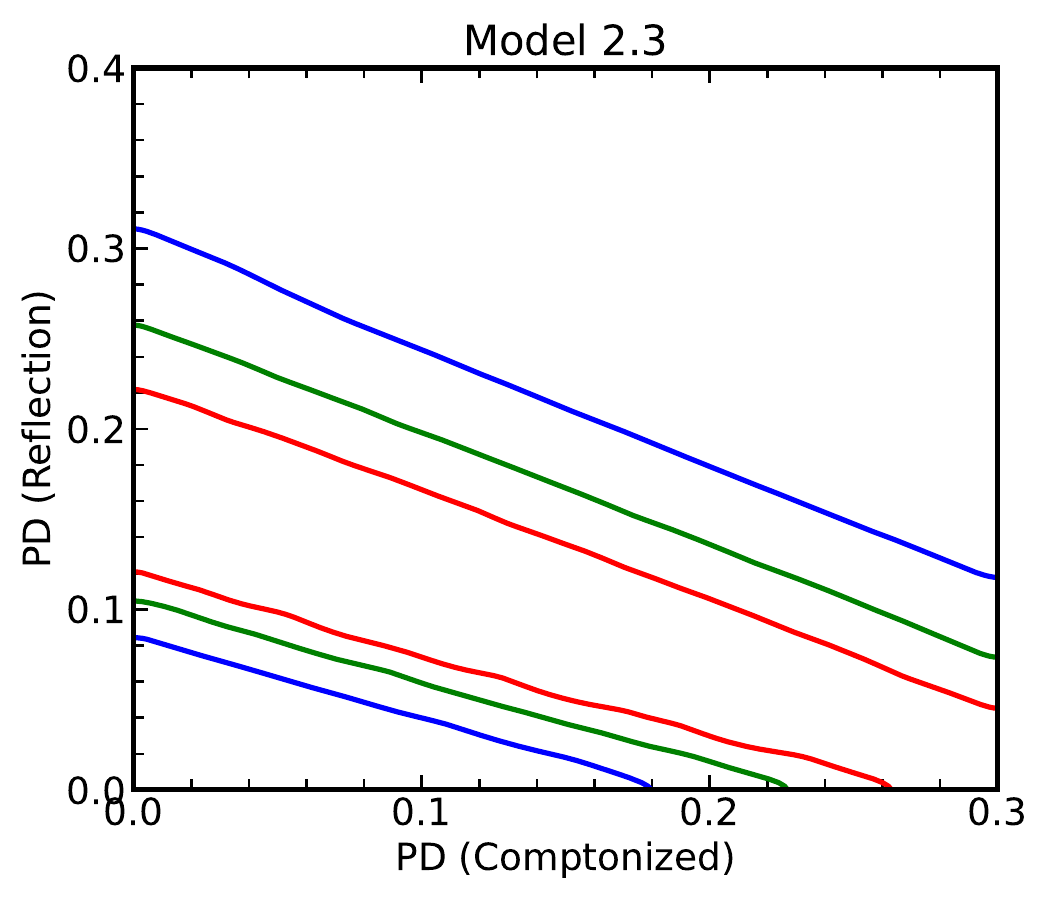}
    \includegraphics[width=0.45\textwidth]{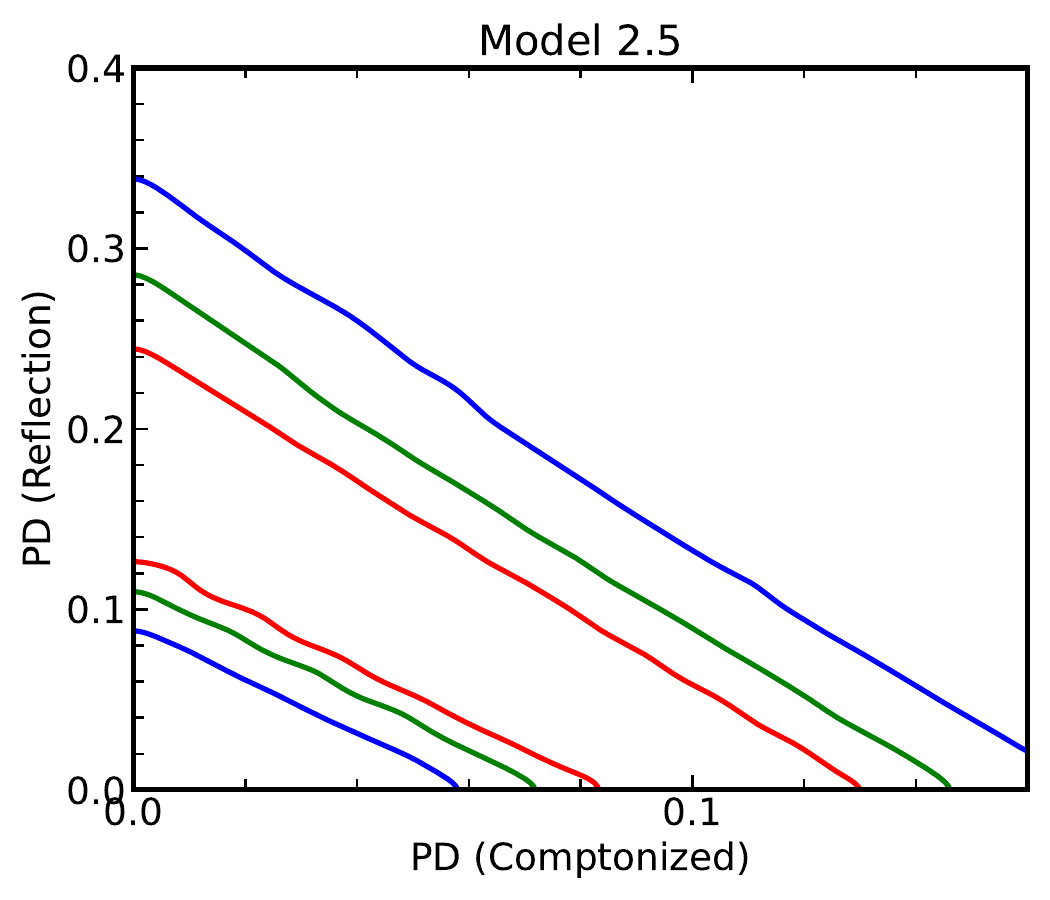}
    \caption{
    (Left) Degeneracy between the PD for the reflection and Comptonization components of Model 2.3. Contours are shown at 68\%, 90\% and 99\% confidence levels. (Right) The same degeneracy as in the left but for model 2.5.}
    \label{ref_cor}
\end{figure*}

\textit{Configuration 1 - } PA$_{\rm comp}-90^{\circ}$ = PA$_{\rm refl}-90^{\circ}$ = PA$_{\rm disk}$. This model provides a $\chi^2/\nu=1085.0/1175$. The best-fit parameters and uncertainties for this configuration can be found in Tab.~\ref{best-fit}. Compared to Model 1, one clear difference is that the full reflection model contributes a flux ($\sim 1 \times 10^{-9}$~erg~s$^{-1}$~cm$^{-2}$) nearly ten times stronger than that of a simple gaussian line ($\sim 9 \times 10^{-11}$~erg~s$^{-1}$~cm$^{-2}$). This is also shown in Fig.~\ref{eemod_delchis}. The PA$_{\rm disk}$ we obtain is consistent with being perpendicular to the total PA and the jet position angle. The PD of the reflection and the Comptonization components are poorly constrained, as shown by the left contour in Fig.~\ref{ref_cor}. This is because the two components are similarly shaped and contribute comparable flux in the 2--8~keV band (see Fig.~\ref{eemod_delchis}), causing the strong degeneracy between the PDs. If we fix the PD$_{\rm comp}$ at 0.5\%, as suggested by calculations in \cite{Farinelli2024A&A...684A..62F}, the PD$_{\rm refl}$ should be 13\%--26\% to explain the data.

\textit{Configuration 2 - } PA$_{\rm comp}$ = PA$_{\rm refl}-90^{\circ}$ = PA$_{\rm disk}$. In this case, the fitting requires the PA of the disk thermal emission to align with the jet position angle, which is thought to be perpendicular to the disk plane. As a consequence, the PA of the reflection component is required to be parallel to the disk plane. Such a configuration is inconsistent with an optically thick accretion disk. Therefore, we do not present the parameters of this fitting. We also note that, although there are a few \ixpe{} observations of the disk-dominated soft state of BH XRBs, there is still a lack of measurements of their radio jet position angle \citep[see][for a recent review]{Dovvciak2024Galax..12...54D}. Therefore, it has not yet been fully confirmed observationally that the PA of the disk thermal emission should be perpendicular to the jet. We should test additional PA configurations if PA$_{\rm disk}$ is not perpendicular to the jet (e.g., PA$_{\rm comp}$ = PA$_{\rm disk}$ = PA$_{\rm refl}$).

\section{Impact of reflection models}
\label{relxillns}

In addition to the \texttt{reflionx} code, another widely used local reflection model is \texttt{xillver} \citep{Garcia2010}. It is included in the relativistic reflection model package: \texttt{relxill}\footnote{\url{https://www.sternwarte.uni-erlangen.de/~dauser/research/relxill/}} \citep{Garcia2013}. To test the impact of reflection models on our conclusion, we fit the spectra with two models:

\begin{itemize}
    \item Model 2.4 - \texttt{mbpo*tbabs*(polconst * diskbb + polconst * comptt + polconst * relconv * xillverNS)}
    \item Model 2.5 - \texttt{mbpo*tbabs*(polconst * diskbb + polconst * comptt + polconst * relxillNS)}
\end{itemize}

We set PA$_{\rm comp}$ = PA$_{\rm refl}$ = PA$_{\rm disk}-90^{\circ}$. The first model is similar to Model 2.3, but replacing \texttt{reflionx\_bb} with \texttt{xillverNS} (from \texttt{relxill} version 2.3). \texttt{xillverNS} is a local reflection model calculated based on the \texttt{xillver} code, assuming the incident spectrum to be a single-temperature blackbody \citep{Garcia2022ApJ...926...13G}. Compared to \texttt{reflionx\_bb}, one improvement of \texttt{xillverNS} is that the disk electron density ($\log(n_{\rm e})$) is a free parameter. In the second model, instead of using the relativistic kernel \texttt{relconv}, we use the self-consistently calculated relativistic reflection model \texttt{relxillNS}, which takes into account the light bending effect more properly \citep{Garcia2014, Liu2025MNRAS.536.2594L, Huang2025ApJ...989..168H}. In both models, the reflection fraction parameter is set to be $-1$. 

\begin{figure*}
    \centering
    \includegraphics[width=0.6\textwidth]{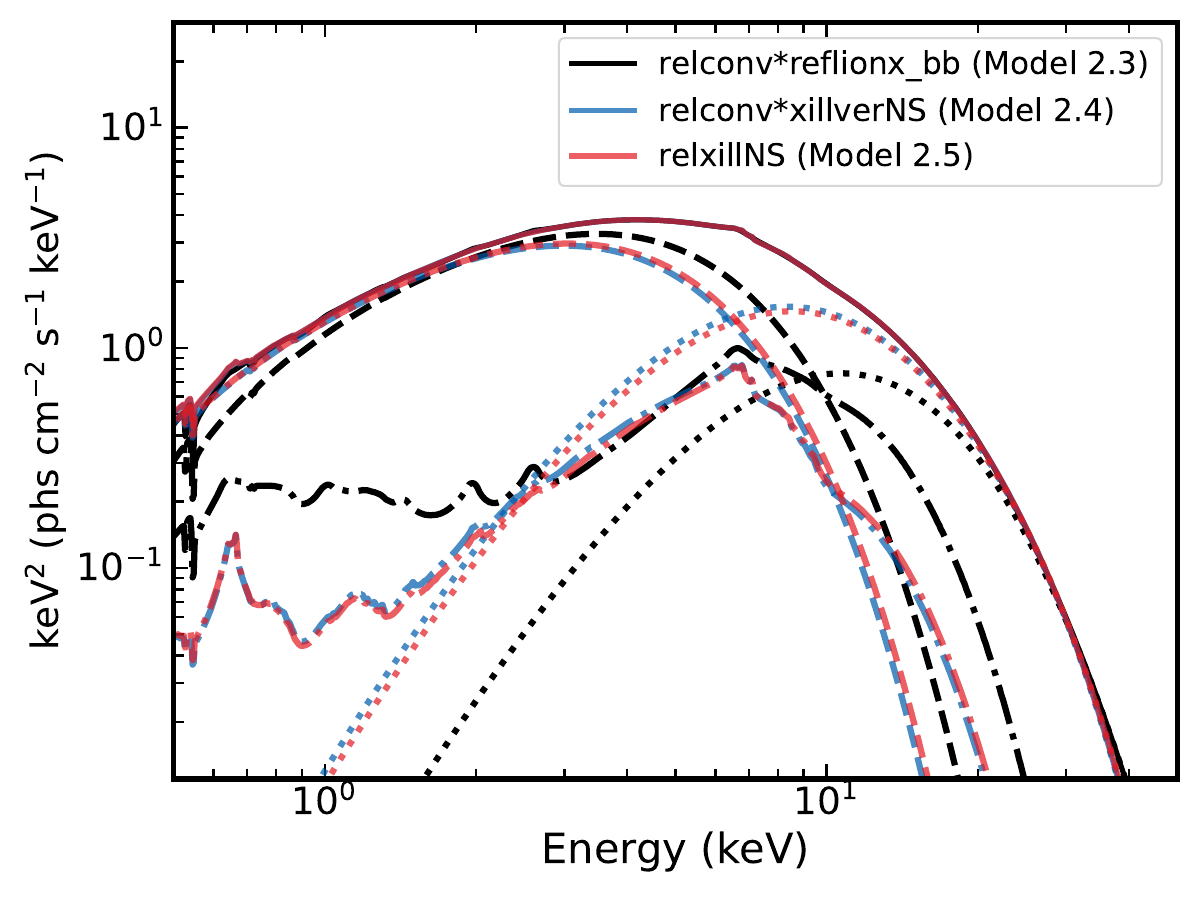}
    \caption{
    Comparison of spectral components for fittings based on different reflection models. The total model is shown as solid lines, while the \texttt{diskbb} and \texttt{comptt} components are represented by dashed and dotted lines, respectively. The reflection component is plotted using dash-dotted lines.
    }
    \label{compare_reflection}
\end{figure*}

The best-fit parameters for the two models are shown in Tab.~\ref{best-fit}. The difference between the two models is minor. They give similar parameters and almost identical spectral shape for each components (see Fig.~\ref{compare_reflection}). 
If we compare the spectral shape in Fig.~\ref{compare_reflection}, the reflection component of Model 2.3 contributes more flux in the bands below 3~keV and above 7~keV. As a consequence, the Comptonization component gets weaker to compensate.


As shown in the right panel of Fig.~\ref{ref_cor}, the \texttt{relxillNS} model exhibits a similar degeneracy between the PD of the reflection and Comptonization components. Since the PD of the Comptonization component should not exceed 1.5\% \citep{Bobrikova2025A&A...696A.181B}, the reflection component must play a significant role in explaining the polarization signal. If fixing the PD$_{\rm comp}$ at 0.5\% \citep{Farinelli2024A&A...684A..62F}, the PD$_{\rm refl}$ should be between 14\%--30\% (90\% confidence level) to explain the data. The conclusion is not changed by the choice of reflection models.


\begin{table*}
    \centering
    \caption{Best-fit parameters and uncertainties}
    \label{best-fit}
    \begin{tabular}{lccccccc}
        \hline\hline
        Component & Parameter & Units & Model 1 & Model 2.3 & Model 2.4 & Model 2.5 \\
        \hline
        \texttt{tbabs} & $N_{\rm H}$ & 10$^{22}$~cm$^{-2}$ & $0.084_{-0.014}^{+0.014}$ & $0.107_{-0.012}^{+0.013}$ & $0.054_{-0.012}^{+0.015}$ & $0.052_{-0.013}^{+0.017}$ \\
        \hline
        \texttt{diskbb}  & $T_{\rm in}$ & keV & $0.87_{-0.06}^{+0.06}$ & $1.458_{-0.03}^{+0.018}$ & $1.26_{-0.11}^{+0.06}$ & $1.28_{-0.11}^{+0.05}$  \\
        \hline
        \texttt{comptt}  & $T_{0}$ & keV & $1.12_{-0.04}^{+0.04}$ & $2.6_{-0.5}^{+0.2}$  & $1.77_{-0.15}^{+0.14}$ &  $1.82_{-0.13}^{+0.22}$  \\
                         & $kT$ & keV & $3.16_{-0.04}^{+0.04}$ & $4.5_{-1.2}^{+10}$ & $3.30_{-0.13}^{+0.2}$ & $3.35_{-0.19}^{+0.21}$ \\
                         & $\tau$ & & $3.89_{-0.09}^{+0.1}$  & $2.0_{-0.9}^{+3}$ & $4.1_{-0.5}^{+0.4}$ & $4.0_{-0.6}^{+0.6}$  \\
        \hline
        \texttt{gaussian} & $E_{\rm line}$ & keV & $6.46_{-0.06}^{+0.05}$ &  & & \\
                          & $\sigma$ & keV & $0.70_{-0.09}^{+0.09}$ & & &  \\
        \hline
        \texttt{reflection} &  & & & \texttt{reflionx} & \texttt{xillverNS} & \texttt{relxillNS} \\
                            & $q$ & & & $3^*$ & $1.5_{-P}^{+0.4}$ & $1.6_{-0.6}^{+0.4}$ \\
                            & $i$ & deg & & $<10$  & $20_{-8}^{+10}$ & $20_{-9}^{+8}$ \\
                            & $R_{\rm in}$ & $R_{\rm ISCO}$ & & $4.9_{-0.7}^{+1.0}$ & $<5$ & $<5$ \\
                            & $\log(\xi)$ & ${\rm erg~cm~s^{-1}}$ & & $2.91_{-0.06}^{+0.13}$ & $3.44_{-0.18}^{+0.13}$ & $3.43_{-0.09}^{+0.13}$ \\
                            & $A_{\rm Fe}$ & solar & & $0.29_{-0.05}^{+0.13}$ & $2.5_{-0.8}^{+0.8}$ & $2.5_{-0.7}^{+0.8}$ \\
                            & $kT_{\rm bb}$ & keV & & $2.13_{-0.11}^{+0.5}$ & $1.83_{-0.21}^{+0.11}$ & $1.88_{-0.06}^{+0.12}$ \\
                            & $\log(n_{\rm e})$ & & & $-$ & $17.4_{-2.1}^{+0.8}$  &  $17.6_{-2.0}^{+0.7}$\\
                            
        \hline
        \texttt{polconst} & PD$_{\rm total}$ & $\%$ & $1.4_{-0.4}^{+0.4}$ & $-$ & & \\
                          & PA$_{\rm total}$ & deg & $136_{-8}^{+8}$ & $-$ & & \\
                          & PD$_{\rm disk}$ & $\%$ &  & $<0.9$ & $<1$ & $<1$ \\
                          & PA$_{\rm disk}$ & deg &  & $45_{-7}^{+7}$  & $46_{-7}^{+7}$ & $46_{-7}^{+7}$ \\
                          & PD$_{\rm comp}$ & $\%$ &  & $<54$ & $<12$ & $<13$ \\
                          & PA$_{\rm comp}$ & deg &  & $=$PA$_{\rm disk}-90$  & $=$PA$_{\rm disk}-90$ & $=$PA$_{\rm disk}-90$ \\
                          & PD$_{\rm refl}$ & $\%$ &  & $<23$ & $<23$ & $<24$ \\
                          & PA$_{\rm refl}$ & deg &  & $=$PA$_{\rm disk}-90$  & $=$PA$_{\rm disk}-90$ & $=$PA$_{\rm disk}-90$ \\
        \hline
        Flux  & Disk & $10^{-9}$~erg & $2.6_{-0.3}^{+0.4}$ & $6.14_{-0.22}^{+0.12}$ & $5.1_{-0.6}^{+0.5}$ & $5.2_{-0.7}^{+0.5}$\\
         (2--8 keV)               & Compton & ~s$^{-1}$~cm$^{-2}$ & $5.1_{-0.4}^{+0.3}$ & $0.55_{-0.09}^{+0.11}$ & $1.68_{-0.14}^{+0.29}$ & $1.57_{-0.22}^{+0.4}$ \\
                        & Gaussian & & $0.087_{-0.014}^{+0.016}$ &  & & \\
                        & Reflection & &  & $1.10_{-0.14}^{+0.23}$ & $1.00_{-0.1}^{+0.24}$ & $0.96_{-0.19}^{+0.28}$ \\
        
        \hline
        Photon flux  & $F_{\rm disk}^{\rm ph}/F_{\rm total}^{\rm ph}$ & \%  & 40.4 & 83.3 & 72.3 & 74.0 \\
             ratio                & $F_{\rm comp}^{\rm ph}/F_{\rm total}^{\rm ph}$ & \% & 59.0 & 5.2 & 16.8 & 15.7 \\
                (2--8 keV)                & $F_{\rm gaus}^{\rm ph}/F_{\rm total}^{\rm ph}$ & \% & 0.6 &  & & \\
                                & $F_{\rm refl}^{\rm ph}/F_{\rm total}^{\rm ph}$ & \% &  & 11.5 & 10.9 & 10.3 \\
        \hline
        $\chi^2/\nu$ & & & 1281.2/1180 & 1085.0/1175  & 1081.6/1173  & 1081.9/1173 \\
        \hline\hline
    \end{tabular} \\
    \textit{Note.} The parameters for the \texttt{mbpo} model and the gain fit parameters are shown in Tab.~\ref{mbpo-para}. The flux in the 2--8 keV band for each component is obtained by applying the \texttt{cflux} model and refitting the data. Uncertainties are given at 90\% confidence level for one parameter of interest.
\end{table*}

\section{Discussion}
\label{sec:discussion}

\subsection{Importance of the reflection component}
\label{sec:importance_of_reflection}

In this work, we reanalyzed the first \ixpe{} observations of the low-mass XRB \src{}, incorporating the quasi-simultaneous \nustar{}, \nicer{} and \integral{} observations. After including the full reflection model, we find that the PA of the disk thermal emission is consistent with being perpendicular to the overall PA and the position angle of the radio jet, in agreement with the findings of \cite{Farinelli2023MNRAS.519.3681F}. This reconfirms that the X-ray polarization properties of \src{} are due to the Comptonization or reflection components (or a combination of both). 

With the full reflection model, the reflection is still not the dominate spectral compoent. It contributes $\sim$ 10\% of the photon flux in the 2--8 keV band, while the disk thermal and the Comptonization components contribute $\sim$ 74\% and 16\%, respectively (Model 2.5). The PD of the disk component should be below 1\%. With the data alone, it is not possible to tightly constrain the PD of the reflection and Comptonization components. We obatin upper limits of 24\% and 13\% respectively.

Ths is due to the strong degeneracy between the PD of the two components (Fig.~\ref{ref_cor}). Such a degeneracy has also been found in other Z sources, such as Sco X-1 \citep{LaMonaca2024ApJ...960L..11L} and GX 340+0 \citep{LaMonaca2024A&A...691A.253L}. From Fig.~\ref{ref_cor}, we observe that the PD of the Comptonized component should be greater than 10\% if the PD of the reflection component is 0. Such a high PD is highly unlikely for the boundary layer emission {\citep[see, e.g.,][]{Gnarini2022MNRAS.514.2561G, Farinelli2024A&A...684A..62F, Bobrikova2025A&A...696A.181B}.} Therefore, the reflection component must be contributing to the polarization signal of the source. The PD of the spreading layer should not exceed 1.5\%, regardless of its geometry \citep{Bobrikova2025A&A...696A.181B}. Based on calculations by \cite{Farinelli2024A&A...684A..62F} using typical parameters for Comptonized emission in the soft state, the boundary layer emission of \src{} should have a PD of up to 0.5\%. In this case, the PD of the reflection component is expected to be between 13\% and 30\%, independent of the specific reflection model used. Such a PD range is feasible for reprocessed emission \citep{Matt1993MNRAS.260..663M}. To demonstrate better this point, we run the \texttt{kynstokes} model \citep{Podgorny2023MNRAS.524.3853P}, which self-consistently calculates the spectral and polarimetric properties of the disk reflection component. We input parameters of \src{} and show PD versus energy for the reflection component in Fig.~\ref{PD_kynstokes}. \texttt{kynstokes} predicts a PD$_{\rm refl}$ exceeding 20\% if the inclination angle is high (e.g., $>60^{\circ}$). This might be in conflict with the low inclination angle we obtained with the \texttt{reflionx\_bb} model. We discuss this tension in Sec.~\ref{sec:spectral_para}. From Fig.~\ref{PD_kynstokes}, we also see that the PD does not show strong variations with energy. Therefore the simple \texttt{polconst} model might be a good approximation at the zeroth-order\footnote{This may not be true for more general cases, especially when the disk is weakly or moderately ionized.}. On top of that, there are some subtle features, such as dips of PD due to depolarized fluorescent lines.


We conducted a further test by directly applying the \texttt{kynstokes} model to the data (see details in Sec.~\ref{sec:kynstokes}). However, this does not yield a good fit, with a $\Delta\chi^2$ of 200 compared to Model 2.3 and an extremely high PD ($\sim$ 20\%) for the Comptonization component, which would be difficult to explain. This is likely because the coronal geometries in kynstokes are too simple to describe the BL/SL regions. The other reason could be that the parameter space of \texttt{kynstokes} is primarily designed for AGNs, assuming the incident spectrum to be a power-law with a high-energy cutoff fixed at 251 keV. We note that Fig.~\ref{PD_kynstokes} remains useful in this case, since the general picture -- that a high-inclination disk results in a highly polarized reflection component -- should not change with the incident spectrum (although it is yet to be tested (Podgorny et al. in prep)).

It should be noted that even the \texttt{reflionx\_bb} model is not entirely self-consistent in this case, because the realistic incident spectrum should be in the form of unsaturated Comptonization (e.g., \texttt{comptt} with low temperature and high optical depth, as obtained in Tab.~\ref{best-fit}). A single-temperature blackbody could mimic such a component and is the best model available. A more self-consistent reflection model, incorporating the correct incident spectrum and including polarization calculations, would enable more detailed characterization of the system. Nevertheless, our basic conclusion --- that the reflection component contributes to the X-ray polarization signal --- remains unaffected, as it depends primarily on the relative photon flux contribution of each component within the IXPE band.

There are other models that can be used to simultaneously fit the spectral and polarimetric properties of the Comptonization component, such as the MONK \citep{Zhang2019ApJ...875..148Z} or KerrC \citep{Krawczynski2022ApJ...934....4K} codes. These models are flexible and can be used to explore various possible geometries of the Comptonizing regions \citep[e.g.,][]{Gnarini2022MNRAS.514.2561G}. To fit observational data, one must use the codes to precompute a grid of models, which can be time-consuming \citep[e.g.,][]{Krawczynski2022ApJ...934....4K}. Therefore, we leave such an analysis for future work. 

We also note that scattering by accretion disk winds could affect the X-ray polarization properties \citep[e.g.,][]{Tomaru2024MNRAS.527.7047T, Nitindala2025A&A...694A.230N}. The resulting PA and PD depend on the illuminating source, as well as the opening angle and optical depth of the disk wind. In the case of \src{}, the Comptonization component from the boundary layer can be approximated as a unpolarized central source, while the disk thermal emission originates from an electron-scattering-dominated disk. Scattering of the two components produces opposite polarization at the inclination angle of \src{} (see Fig. 5 and Fig. 8 of \citealt{Nitindala2025A&A...694A.230N}). The outcome and its impact on the required PD of the reflection component is unclear. A quantitative understanding of this effect requires more detailed simulations. In the same spirit, scattering by the accretion disk corona (e.g., hot plasma above the disk) can also produce a polarization signal perpendicular to the disk plane. Such an accretion disk corona structure has been suggested to exist in \src{} due to the detection of highly ionized emission lines. \citep[e.g.][]{Schulz2009ApJ...692L..80S, Mizumoto2025arXiv251200589M}.

\subsection{Inclination angle}
\label{sec:spectral_para}

As shown in Tab.~\ref{best-fit}, Model 2.3 gives a low inclination angle for the accretion disk, with an upper limit of 10 degree. This value is significantly lower than the orbital inclination angle ($62.5^{\circ} \pm 4^{\circ}$, \citealt{Orosz1999MNRAS.305..132O}), although the two angles are not necessarily identical. There are additional indications of a high inclination angle for \src{}, such as the dips observed in its light curves \citep{Balucinska-Church2011A&A...530A.102B}. As a prototype of Cyg-like sources, it has also been suggested to have a higher inclination than Sco-like sources \citep{Kuulkers1995A&A...303..801K}. As discussed in Sec.~\ref{sec:importance_of_reflection}, the \ixpe{} data also require a relatively high inclination angle for the disk. Otherwise, the reflection component would not produce sufficient polarization to match the observations. Therefore, a face-on configuration with a low-inclination disk is disfavored.

In the literature, there is already some tension for the measurements of its disk inclination angle with the reflection method. With the 2015 \nustar{} data and \texttt{reflionx\_bb} model, \cite{Mondal2018MNRAS.474.2064M} measured an upper limit of $\sim 20^{\circ}$. While \cite{Ludlam2022ApJ...927..112L} found an inclination of $67^{\circ} \pm 4^{\circ}$ with the 2019 \nustar{} data and the \texttt{relxillNS} \citep{Garcia2022ApJ...926...13G} model. Therefore, we conduct two more fittings by replacing the \texttt{relconv*reflionx\_bb} component in Model 2.3 with \texttt{relconv*xillverNS} and \texttt{relxillNS} (see details in Sec.~\ref{relxillns}). These models indeed provide a higher inclination angle ($10^{\circ}$--$40^{\circ}$) without affecting the conclusion that a highly-polarized reflection component is required to explain the polarization signal of \src{}.

We also suggest to take the inclination angle measurement with the reflection models in this work with caution. The reflection spectra are only sensitive to the inner region of the accretion disk, which may have an inclination angle differing from the orbital one. The reflection method can also be affected by systematic uncertainties, such as the unknown or the simplified modeling of the primary source illuminating the disk \citep{Bambi2021}. In some cases, the systematic uncertainty
can be $\sim$~30$^{\circ}$ \citep[e.g.,][]{Garcia2018ApJ86425G, Connors2022}.


\subsection{Comparison to other sources}

\ixpe{} has observed around 10 weakly magnetized NS LMXBs, including both atoll and Z sources \citep[see][for a recent review]{Ursini2024Galax..12...43U}. Before this work, spectro-polarimetric analyses including full reflection models have been applied to \textit{IXPE} observations of a few NS LMXBs where strong reflection features are shown: Sco~X-1 \citep{LaMonaca2024ApJ...960L..11L}, GX~5--1 \citep{Fabiani2024A&A...684A.137F}, GX~9+9 \citep{Ursini2023A&A...676A..20U}, 4U~1624--49 \citep{Gnarini2024A&A...690A.230G}, GX~340+0 \citep{LaMonaca2024A&A...691A.253L} and GX~349+2 \citep{LaMonaca2025A&A...702A..40L}. As in the case of \src{}, there is usually a strong degeneracy between the PD of the reflection and Comptonization components \citep{LaMonaca2024ApJ...960L..11L, LaMonaca2024A&A...691A.253L}. If assuming the Comptonized component is not polarized, a high PD of 10-30\% for the reflection component is required by the data. There are indications that these sources have relatively high inclination angle \citep[$>40^{\circ}$,][]{Frank1987A&A...178..137F, Iaria2020A&A...635A.209I, Fomalont2001ApJ...553L..27F, Homan2018ApJ...853..157H}, which is necessary to produce a highly polarized reflection component.

Among these sources, Sco~X-1 shows peculiar properties. The PA of its full band or any single spectral component measured with \textit{IXPE} is not aligned with previous measurements by OSO-8 \citep{Long1979ApJ...232L.107L}, PolarLight \citep{Long2022ApJ...924L..13L}, or the previously measured jet position angle \citep{Fomalont2001ApJ...553L..27F}. The PA variations may indicate changes in the coronal geometry, but we also suggest that the precession of the inner accretion disk and the corresponding changes in the PA of the reflection component might be a possibility. The precession can be caused by a misalignment between the disk and the NS spin axis. Such a scenario has also been suggested to explain the energy-dependent PA of GX~5--1 \citep{Fabiani2024A&A...684A.137F}. Another possible explanation for the PA variation is Faraday rotation caused by coherent magnetic fields in the photosphere \citep[e.g.,][]{Barnier2024ApJ...977..201B}. In the case of a vertical magnetic field, a constant rotation that depends linearly on the field strength is expected. Therefore, a change in the field strength could cause the PA variation. However, for azimuthal or radial magnetic field configurations, depolarization is expected rather than a PA variation. A similar time dependence of the PA has also been observed in two other sources: GX 13+1 \citep{Bobrikova2024A&A...688A.170B, Bobrikova2024A&A...688A.217B, DiMarco2025ApJ...979L..47D} and Cir X-1 \citep{Rankin2024ApJ...961L...8R}. In GX 13+1, the PA variation is associated with dips \citep{DiMarco2025ApJ...979L..47D}, whereas in Cir X-1, it is related to the orbital variation \citep{Rankin2024ApJ...961L...8R}. In an axisymmetric system, the PA should be either parallel or orthogonal to the symmetry axis. However, the PA variation is not 90$^{\circ}$ in either Cir X-1 or GX 13+1. This may indicate an asymmetry in the system, such as a misalignment between the accretion disk and the neutron star's spin axis \citep{Rankin2024ApJ...961L...8R, Bobrikova2024A&A...688A.170B}. Another way to break the symmetry could be partial obscuration of the extended accretion disk corona by clumps \citep[e.g.,][]{DiMarco2025ApJ...979L..47D}.

A general trend has been observed where Z-sources (highly accreting) exhibit a higher PD in the 2--8~keV band compared to atoll sources (slowly accreting). This trend may be explained by the reflection scenario: when the accretion rate is high, the spreading layer extends to higher latitudes, thereby better illuminating the accretion disk and producing a stronger reflection component. This scenario should be tested with spectro-polarimetric analyses of more sources covering a variety of accretion states.

Another trend is that the PD tends to increase with energy, which is apparent for both atoll and Z sources. The trend suggests that the harder emission components, such as the Comptonization and reflection, are dominating the polarization signal. In particular, relatively high PDs have been found in the harder band of a few sources: $10.3 \pm 2.4$\% in the 7--8 keV band for 4U 1820--303 \citep{DiMarco2023ApJ...953L..22D}, $6 \pm 2$\% in the 6--8 keV band for 4U 1624--49 \citep{Saade2024ApJ...963..133S} and $5.4 \pm 0.7$\% in the 5--8 keV band for GX 5--1 \citep{Fabiani2024A&A...684A.137F}. These high PDs are difficult to explain with the Comptonization component alone, suggesting an important role of the reflection component.

Within Z-sources, there is a subtle trend that the HB (with a harder spectrum) appears to show a higher PD than those in the NB and FB (with a softer spectrum). This has been found in several sources: XTE J1701--462 \citep{Cocchi2023A&A...674L..10C}, GX 5--1 \citep{Fabiani2024A&A...684A.137F} and GX 340+0 \citep{LaMonaca2024A&A...691A.253L, LaMonaca2025A&A...702A.101L}. In other sources that \ixpe{} does not cover all branches, similar values of PD have been observed in the same states \citep[e.g.,][]{Lavanya2025ApJ...985..229L, LaMonaca2025A&A...702A..40L}, i.e. $\sim$~4\% for the HB and $\sim$~2\% for the NB \citep[see Table 2 of][]{Gnarini2025A&A...699A.230G}. We also note that a similar trend has been found in BH XRBs systems: the PD was observed to significantly decrease from hard/intermediate states, where reflection should be strong, to the soft state, where the disk thermal emission dominates the spectrum \citep[e.g.,][]{Veledina2023ApJ...958L..16V, Ingram2024ApJ...968...76I, Svoboda2024ApJ...966L..35S, Podgorny2024A&A...686L..12P}. It is worth noting that the reflection component could still be important even in the disk-dominated soft state. This is illustrated by Cygnus~X-1, where the disk reflection of the returning radiation is key to explain its X-ray polarization properties \citep{Steiner2024ApJ...969L..30S}. The PA is generally consistent along the Z-track, although tentative variations have been suggested for Sco~X-1 and GX~349+2 \citep[see Figure~7 of][]{Gnarini2025A&A...699A.230G}. It should be noted that such variations are not statistically significant with the current data \citep[e.g.,][]{Kashyap2025ApJ...986..207K}.


\section{Conclusion}

In this work, we conducted a spectro-polarimetric analysis of \src, using data from \ixpe{}, \nustar{}, \nicer{}, and \integral{}. To test the hypothesis that the source polarization signal is dominated by the reflection component, as proposed by \cite{Farinelli2024A&A...684A..62F}, we model reflection utilizing a full and rigorous reflection model rather than approximating this effect as just a Fe-K fluorescence line. We find a strong degeneracy between the PD of the reflection and Comptonization components. If the Comptonization component alone were responsible for the X-ray polarization signal, it would require an unreasonably high PD. Therefore, it is highly likely that the reflection component contributes significantly to the observed polarization. Given a typical PD for the Comptonization component, a PD of 13-30\% for the reflection component can well explain the \ixpe{} data. 


\begin{acknowledgments}
{\it Acknowledgments:}
We thank Wenda Zhang, Menglei Zhou, Pengju Wang and Long Ji for insightful discussions. This work was supported by the National Natural Science Foundation of China (NSFC), Grant No. 12250610185 and 12261131497, and the Natural Science Foundation of Shanghai, Grant No. 22ZR1403400. AI acknowledges support from the Royal Society. AJY acknowledges this work was supported by the Science and Technology Facilities Council grant number ST/Y001990/1. J.P. thanks the institutional support from RVO:67985815.
\end{acknowledgments}

\bibliography{bibtex}{}
\bibliographystyle{aasjournal}

\appendix

For clarity, we do not show the parameters of the \texttt{mbpo} model in Tab.~\ref{best-fit}, as they simply fit the cross-calibration between instruments. These parameters are provided in Tab.~\ref{mbpo-para}.

\begin{table*}
    \centering
    \caption{Best-fit parameters and uncertainties of the \texttt{mbpo} model and the gain parameters}
    \label{mbpo-para}
    \begin{tabular}{lcccccc}
        \hline\hline
        Instrument & Parameter & Model 1 & Model 2.3 & Model 2.4 & Model 2.5 \\
        \hline\hline
        \texttt{mbpo} & & & & & \\
        \hline
        \nustar{}-FPMA & $\Delta\Gamma$ & $0^*$   & $0^*$ & $0^*$ & $0^*$ \\
                        & Norm & $1^*$ & $1^*$ & $1^*$ & $1^*$ \\
        \nustar{}-FPMB & $\Delta\Gamma$ & $0.002_{-0.004}^{+0.004}$ & $0.002_{-0.004}^{+0.004}$ & $0.002_{-0.004}^{+0.004}$ & $0.002_{-0.004}^{+0.004}$ \\
                        & Norm & $1.0030_{-0.0026}^{+0.0026}$ & $1.0029_{-0.0025}^{+0.0025}$  & $1.0029_{-0.0026}^{+0.0026}$ & $1.0029_{-0.0026}^{+0.0026}$ \\
        \hline
        \nicer{} (April 30) & $\Delta\Gamma$ & $0.107_{-0.008}^{+0.008}$ & $0.104_{-0.007}^{+0.009}$  & $0.104_{-0.007}^{+0.008}$ & $0.104_{-0.007}^{+0.007}$ \\
                        & Norm & $1.026_{-0.007}^{+0.007}$ & $1.023_{-0.007}^{+0.008}$    & $1.023_{-0.007}^{+0.007}$ & $1.023_{-0.007}^{+0.007}$  \\
        \nicer{} (May 01) & $\Delta\Gamma$ & $0.141_{-0.008}^{+0.009}$ & $0.138_{-0.008}^{+0.009}$  & $0.138_{-0.008}^{+0.008}$ & $0.137_{-0.008}^{+0.008}$ \\
                        & Norm & $0.971_{-0.003}^{+0.003}$ & $0.970_{-0.003}^{+0.003}$  & $0.970_{-0.003}^{+0.003}$ & $0.970_{-0.003}^{+0.003}$  \\
        \hline
        \integral{}-ISGRI & Norm & $0.7_{-0.3}^{+0.3}$ & $0.64_{-0.29}^{+0.29}$  & $0.64_{-0.29}^{+0.29}$ & $0.64_{-0.29}^{+0.29}$ \\
        \integral{}-JEM X-1 & $\Delta\Gamma$ & $-0.04_{-0.06}^{+0.06}$  & $-0.04_{-0.06}^{+0.06}$   & $-0.04_{-0.06}^{+0.06}$ & $-0.04_{-0.06}^{+0.06}$ \\
                        & Norm & $1.08_{-0.03}^{+0.03}$ & $1.078_{-0.03}^{+0.017}$     & $1.08_{-0.03}^{+0.03}$ &  $1.08_{-0.03}^{+0.03}$\\
        \integral{}-JEM X-2 & $\Delta\Gamma$ & $0.00_{-0.06}^{+0.06}$ & $0.00_{-0.06}^{+0.06}$    & $0.00_{-0.06}^{+0.06}$ & $0.00_{-0.06}^{+0.06}$ \\
                        & Norm & $0.96_{-0.03}^{+0.03}$ & $0.959_{-0.03}^{+0.029}$      & $0.959_{-0.029}^{+0.029}$ & $0.959_{-0.03}^{+0.029}$ \\
        \hline
        \ixpe{}-DU1 & $\Delta\Gamma$ & $-0.22_{-0.04}^{+0.04}$ & $-0.19_{-0.06}^{+0.04}$   & $-0.21_{-0.04}^{+0.04}$ & $-0.21_{-0.04}^{+0.04}$  \\
                    & Norm & $0.902_{-0.026}^{+0.025}$ & $0.923_{-0.028}^{+0.025}$     & $0.912_{-0.026}^{+0.024}$ & $0.912_{-0.025}^{+0.025}$ \\
        \ixpe{}-DU2 & $\Delta\Gamma$ & $-0.18_{-0.06}^{+0.05}$ & $-0.15_{-0.06}^{+0.05}$     & $-0.17_{-0.06}^{+0.05}$ & $-0.17_{-0.06}^{+0.05}$  \\
                    & Norm & $0.94_{-0.04}^{+0.03}$  & $0.96_{-0.04}^{+0.03}$    & $0.94_{-0.03}^{+0.03}$  & $0.95_{-0.04}^{+0.03}$ \\
        \ixpe{}-DU3 & $\Delta\Gamma$ & $-0.10_{-0.04}^{+0.04}$  & $-0.08_{-0.04}^{+0.04}$    & $-0.09_{-0.02}^{+0.04}$ & $-0.09_{-0.04}^{+0.04}$ \\
                    & Norm & $0.945_{-0.027}^{+0.024}$  & $0.965_{-0.029}^{+0.025}$     & $0.954_{-0.026}^{+0.023}$ & $0.954_{-0.025}^{+0.025}$  \\
        \hline\hline
        \texttt{gain shift} & & &  \\
        \ixpe{}-DU1 & slope & $1.152_{-0.01}^{+0.011}$ & $1.149_{-0.011}^{+0.011}$  & $1.150_{-0.005}^{+0.011}$ & $1.086_{-0.009}^{+0.009}$ \\
                    & offset & $-0.187_{-0.014}^{+0.014}$ & $-0.192_{-0.013}^{+0.012}$ & $-0.189_{-0.014}^{+0.014}$ & $-0.094_{-0.012}^{+0.013}$  \\
                    \hline
        \ixpe{}-DU2 & slope & $1.231_{-0.015}^{+0.017}$ & $1.225_{-0.015}^{+0.018}$ & $1.230_{-0.015}^{+0.016}$ & $1.145_{-0.014}^{+0.014}$ \\
                    & offset & $-0.197_{-0.018}^{+0.017}$ & $-0.201_{-0.017}^{+0.018}$ & $-0.201_{-0.018}^{+0.016}$ & $-0.088_{-0.018}^{+0.017}$ \\
        \hline
        \ixpe{}-DU3 & slope & $1.131_{-0.01}^{+0.011}$ & $1.128_{-0.01}^{+0.011}$ & $1.13_{-0.01}^{+0.01}$ & $1.062_{-0.009}^{+0.006}$ \\
                    & offset & $-0.123_{-0.014}^{+0.013}$ & $-0.129_{-0.013}^{+0.011}$ & $-0.126_{-0.014}^{+0.013}$ & $-0.022_{-0.012}^{+0.013}$  \\

        \hline\hline
    \end{tabular} \\
\end{table*}

\section{\texttt{kynstokes} model}
\label{sec:kynstokes}

In this section, we test a recently developed model, \texttt{kynstokes} \citep{Podgorny2023MNRAS.524.3853P}, which can simultaneously fit the spectral and polarimetric properties of the disk reflection component.

This model self-consistently calculates the energy spectrum of the reflection component, as well as the corresponding PA and PD as a function of energy. The incident spectrum for the \texttt{kynstokes} model is a power-law with a high-energy cutoff at 251~keV.  The full model in \texttt{XSPEC} format is: \texttt{mbpo*tbabs*(polconst * diskbb + polconst * comptt + kynstokes)}. We assume a spin of $a_*=0.17$ and test two configurations of the primary source: (1) a lamppost with a height $h$, and (2) modeling the emissivity profile with a power-law, assuming a slab corona geometry.

\begin{figure*}
    \centering
    \includegraphics[width=0.7\textwidth]{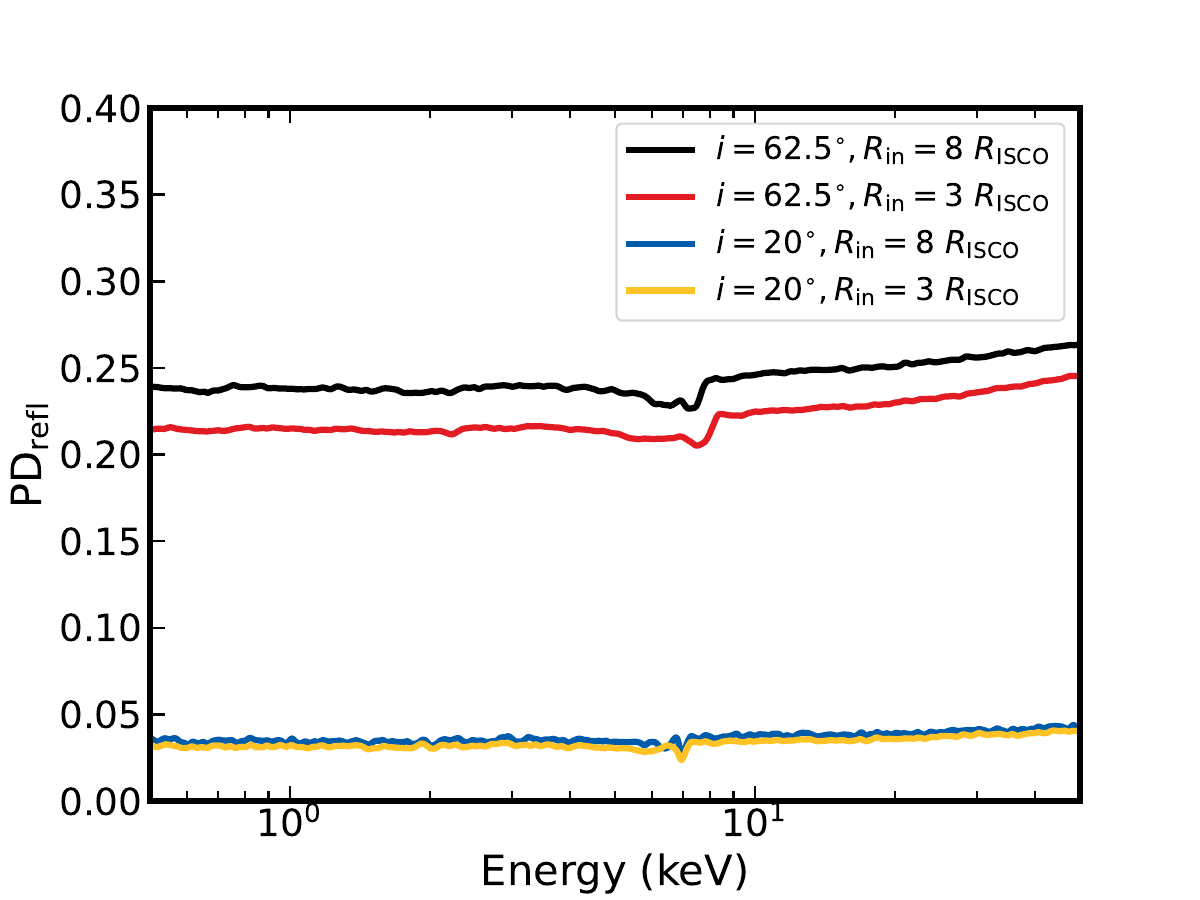}
    \caption{
    The reflected-only PD versus energy in the lamppost scheme derived from the \texttt{kynstokes} model. We set spin to $a_*=0.17$, photon index $\Gamma=2.0$, luminosity of the Comptonization component $L_{\rm 2-10~keV}/L_{\rm Edd}=0.08$, lamppost height $h=5~GM/c^2$. The intrinsic PD of the Comptonization component is 0. We shown cases of two different inclination angles and two inner disk radii.
    }
    \label{PD_kynstokes}
\end{figure*}

The model yields significantly worse fitting statistics ($\Delta\chi^2 \sim 200$) compared to Model 2.3. The main source of the $\Delta\chi^2$ is the energy spectra. The reason is that \src{} requires a soft thermal spectrum to illuminate the disk, whereas the current version of \texttt{kynstokes} assumes a hard power-law spectrum. The model also requires the PD$_{\rm comp}$ to be 15-25\%, which is too high to be realistic \citep[e.g.,][]{Gnarini2022MNRAS.514.2561G}. This might be mitigated by using a more suitable incident spectrum, but we note that other factors should also be considered. For example, a geometry more appropriate than a lamppost for the boundary layer/spreading layer could be essential for correctly calculating the polarization. Therefore, we do not show the best-fit parameters for this model to avoid misleading interpretations. We note that this does not mean the \texttt{kynstokes} model is not useful in this work. The trend shown in Fig.~\ref{PD_kynstokes}, where a higher PD corresponds to a higher inclination angle, should not change with the incident spectrum. Using a blackbody as the incident spectrum is only important for accurately determining the flux of each component, which is crucial for understanding how each component contributes to the polarization signal. Such a model should be straightforward to develop based on \texttt{kynstokes}. The current \texttt{kynstokes} model can still be applied to XRBs in the hard state, where the accretion disk is illuminated by a power-law spectrum.

\end{document}